\documentclass[11pt]{article}
\usepackage{amssymb}
\usepackage{amsmath}
\usepackage[all]{xy}
\input{epsf}
\usepackage{epsfig}
\numberwithin{equation}{section}

\def\be{\begin{equation}}
\def\ee{\end{equation}}
\def\ba{\begin{align}}
\def\ea{\end{align}}
\def\beq{\begin{eqnarray}}
\def\eeq{\end{eqnarray}}

\def\p{\partial}

\def\T{\mathcal{T}}
\def\H{\mathcal{H}}

 \input{epsf}

 \usepackage{epsfig}

\begin{document}

\title{\Large{\bf Fusion of line operators in conformal sigma-models on supergroups, and the Hirota equation
}} 
\author{Raphael Benichou
}
\date{}
\maketitle
\begin{center}
  Theoretische Natuurkunde, Vrije Universiteit Brussel and \\
The International Solvay Institutes,\\
 Pleinlaan 2, B-1050 Brussels, Belgium \\
 \textsl{raphael.benichou@vub.ac.be}
\end{center}

\begin{abstract}
We study line operators in the two-dimensional sigma-model on $PSl(n|n)$ using the current-current OPEs.
We regularize and renormalize these line operators, and compute their fusion up to second order in perturbation theory.
In particular we show that the transfer matrix associated to a one-parameter family of flat connections is free of divergences.
Moreover this transfer matrix satisfies the Hirota equation (which can be rewritten as a Y-system, or Thermodynamic Bethe Ansatz equations) for all values of the two parameters defining the sigma-model.
This provides a first-principles derivation of the Hirota equation which does not rely on the string hypothesis nor on the assumption of quantum integrability.
\end{abstract}

\newpage

\tableofcontents

\newpage


\section{Introduction}

Integrability plays an important r\^ole in the study of the AdS/CFT correspondence \cite{Maldacena:1997re}. 
Recently an infinite set of equations has been proposed to describe the exact spectrum of string theory on $AdS_5 \times S^5$ \cite{Gromov:2009tv}\cite{Gromov:2009bc}.
These equations take the form of a Y-system.
The Thermodynamic Bethe Ansatz equations can be derived from the Y-system once some analytic properties are specified.
Alternatively the Y-system can be written as a T-system, also known as the Hirota equation.
In this paper we focus on the latter form. The Hirota equation is a quadratic relation between commuting scalar operators $\T_R(u)$:
\be\label{Hirota} \T_{a,s}(u + 1)\T_{a,s}(u - 1) = 
  \T_{a+1,s}(u+1)\T_{a-1,s}(u-1) +  \T_{a,s+1}(u-1)\T_{a,s-1}(u+1)\ee
The operator $\T_R(u)$ is understood \cite{Gromov:2009tv}\cite{Gromov:2010vb} as the trace of the monodromy matrix associated to the one-parameter family of flat connections of the string worldsheet theory \cite{Bena:2003wd}.
These matrices can be taken in different representations $R$ of the global symmetry superalgebra. 
The representations appearing in the Hirota equation are labeled by two integer indices $(a,s)$ that take value in a ``T-hook'' lattice.
These representations are described by rectangular Young tableaux with a number of rows and columns respectively related to the indices $a$ and $s$. 
The parameter $u$ is essentially the spectral parameter.

The $AdS_5$/$CFT_4$ Y-system was derived in \cite{Gromov:2009bc} following the Thermodynamic Bethe Ansatz approach \cite{Zamolodchikov:1989cf}.
This derivation relies on the string hypothesis (see \cite{Arutyunov:2009zu}),
namely that all the eigenstates of the (mirror) model that contribute in the thermodynamic limit at large volume are (bounds states of) elementary particles.

T-systems, or the equivalent Y-systems, are rather ubiquitous in the study of integrable models (see \cite{Kuniba:2010ir} for a recent review).
A generic method to compute the spectrum of integrable relativistic two-dimensional sigma-models starting from these equations has been described in \cite{Gromov:2008gj} (see also \cite{Kazakov:2010kf}).

In this paper we consider two-dimensional non-linear sigma-models on the supergroup $PSl(n|n)$.
These models were first studied in \cite{Berkovits:1999im}\cite{Bershadsky:1999hk}. 
The main motivation was that the sigma-model on $PSU(1,1|2)$ is directly relevant for the quantization of string theory on $AdS_3\times S^3$ supported by NSNS and/or RR fluxes.
Later these models have been used to describe (non-supersymmetric) condensed matter systems, for instance the quantum Hall effect \cite{Zirnbauer:1999ua} and disordered fermion systems \cite{Guruswamy:1999hi}.

The sigma-models on $PSl(n|n)$ admit a one-parameter family of flat connections constructed from the currents associated to the global symmetry \cite{Benichou:2010rk}.
Consequently these models are classically integrable: the monodromy matrix associated to the one-parameter family of connections encodes an infinite number of conserved charges.
There is good hope that integrability persists at the quantum level. Indeed the quantum current-current OPEs \cite{Ashok:2009xx}\cite{Benichou:2010rk} are compatible with the Maurer-Cartan equation, which is responsible for the flatness of the connections and thus integrability at the classical level.

In the present paper we take a new step towards the proof of quantum integrability of these models, and their solution.
We study the quantum behavior of the monodromy matrices using the current-current OPEs. 
In particular we show that the trace of the monodromy matrices satisfy the Hirota equation \eqref{Hirota} up to second order in perturbation theory.
This derivation of the Hirota equation is very direct and does not require the assumption of quantum integrability nor the string hypothesis.
Our approach is close in spirit to \cite{Bazhanov:1994ft}, where a T-system was derived for a family of minimal models using a free-field representation of the Virasoro algebra.

Even if conformal symmetry is much more constraining in two dimensions, the $AdS_3$/$CFT_2$ examples of the AdS/CFT correspondence are not better understood than their higher-dimensional counterparts.
One reason is that string theory in $AdS_3$ with RR fluxes is as difficult to study as in higher-dimensional spacetimes.
It is expected that integrability is a powerful tool to study some of these examples (see \cite{Pakman:2009mi} for recent developments).
The results presented in this paper are a new step in that direction.
More generally the tools developed in \cite{Ashok:2009xx}\cite{Benichou:2010rk} allow for a worldsheet approach to quantum string theory in RR backgrounds. For instance in \cite{Benichou:2010rk} the conformal dimensions of the low-level states were computed perturbatively. In \cite{Ashok:2009jw} the target-space super-Virasoro generators were constructed for string theory in $AdS_3 \times S^3 \times T^4$ with RR fluxes, generalizing the results of \cite{Giveon:1998ns} valid in NS backgrounds. 

The sigma-models on supergroups we study share a lot of structure with sigma-models on supercosets%
\footnote{We are considering sigma-models of ``pure-spinor'' type, as opposed to ``Green-Schwarz'' type. They include a kinetic term for the fermions that breaks kappa symmetry.} %
 relevant for the AdS/CFT correspondence in various dimensions \cite{Berkovits:1999zq}\cite{Adam:2007ws}.
The sigma-models on supergroups are slightly simpler to deal with since it is easy to preserve covariance under the full supergroup at all steps of the computations.
In particular it is likely that the computations presented here can be adapted in a straightforward way for the pure-spinor string in $AdS_5\times S^5$ \cite{Berkovits:2000fe}\cite{Bedoya:2010av}.
Some of the first-order computations have already been done in \cite{Mikhailov:2007mr}\cite{Mikhailov:2007eg}, and the pattern of simplifications and cancellations is indeed very similar to the one we observe in this paper.

The plan of the paper is the following.
In section \ref{setup} we introduce the models and summarize the results of \cite{Ashok:2009xx}\cite{Benichou:2010rk} concerning the current-current OPEs. We also describe the one-parameter family of flat connections and the line operators that we will study.
The collisions of the integrated connections in a line operator typically produce divergences in a quantum theory. These divergences imply that a regularization procedure is needed to properly define the quantum line operators. This issue is discussed in section \ref{renormalization}. Here we also show that the trace of the monodromy matrix is free of divergences at least up to second order in perturbation theory.
In section \ref{fusion} we compute the fusion of line operators up to second order in perturbation theory.
Eventually we show in section \ref{proofHirota} that the trace of the monodromy matrix satisfies the Hirota equation \eqref{Hirota}.


\section{Flat connections in conformal sigma-models on supergroups}\label{setup}

In this section we introduce the models under study, and present the quantum properties of the flat connections.


\subsection{Current algebra in conformal sigma models on supergroups}

We consider a two-dimensional non-linear sigma model on the supergroup $PSl(n|n)$. The action is:
\begin{align}\label{action}
&S = S_{kin} + S_{WZ}\cr
&S_{kin} =  \frac{1}{ 16 \pi f^2}\int d^2 z Tr'[- \partial^\mu g^{-1}
\partial_\mu g]
\cr
&S_{WZ} = - \frac{ik}{24 \pi} \int_B d^3 y \epsilon^{\alpha \beta \gamma}
Tr' (g^{-1} \partial_\alpha g g^{-1} \partial_\beta g   g^{-1} \partial_\gamma g )
\end{align}
The model has two parameters $f^2$ and $k$. The former sets the curvature radius of the target space. The latter is an integer. In the special case of $PSU(1,1|2)$, the target space is $AdS_3 \times S^3$ embedded in a superspace with eight fermionic dimensions. Then the integer $k$ gives the amount of NSNS fluxes in this background \cite{Berkovits:1999im}.
The euclidean path-integral is well defined if the parameters satisfy $|k f^2|\le 1$ \cite{Zirnbauer:1999ua}. For $k f^2 = \pm 1$ the model is a WZNW model.

The supergroup $PSl(n|n)$ has a vanishing Killing form. Together with the uniqueness of the rank-three invariant tensor, this ensures that the model is conformal for any choice of the parameters $f^2$, $k$ \cite{Bershadsky:1999hk}\cite{Berkovits:1999im}. Notice that the supergroups $OSp(2n+2|2n)$ also have vanishing Killing form, and it is likely that our results also apply to these supergroups.
The vanishing of the Killing form implies that the double contraction of structure constants of the superalgebra vanishes:
\be f^{acd} {f^b}_{cd} = 0 \ee
This identity plays a central role in the computations presented in this paper.
It means that the dual Coxeter number of the supergroup vanishes.
Notice that the superalgebra contains bosonic and fermionic generators. We will not keep track of the signs that appear in the computations because of the fermionic nature of some generators. They can be consistently restored at each step of the computations.

The model has a global symmetry $G_L \times G_R$ associated to the left- and right- multiplication by a group element.
In the following we focus on the currents associated to the left symmetry, but everything can equivalently be written in terms of the currents associated to the right symmetry.

For our purposes it is convenient to normalize%
\footnote{The normalization of the currents differs from the one used in \cite{Ashok:2009xx}\cite{Benichou:2010rk} by a factor of $f^2$.} %
 the left-currents as:
\begin{align} & j_{z} = -\eta \p g g^{-1} \cr
 & j_{\bar z} = -(1-\eta) \bar \p g g^{-1} \end{align}
with:
\be \eta = \frac{1 + kf^2}{2} \ee
The dimensionless parameter $\eta$ takes values between zero and one.
For $\eta=1$ or $\eta=0$ the model reduces to a WZNW model.
The current is conserved:
\be\label{CC} \bar \p j^a_z + \p j_{\bar z}^a = 0 \ee
and it satisfies the Maurer-Cartan equation:
\be\label{MC} (1-\eta) \bar \p j^a_z - \eta \p j^a_{\bar z} + i {f^a}_{bc} : j^c_z j^b_{\bar z}: = 0 \ee
%
The Maurer-Cartan equation holds classically. There is good hope that it also holds exactly as an operator equality in the quantum theory\footnote{Since the Maurer-Cartan equation is closely related to the integrability properties of the model, the results derived in section \ref{proofHirota} give further evidence in favor of this conjecture.}
\cite{Benichou:2010rk}.

The current-current OPEs read \cite{Ashok:2009xx}:
\begin{align}\label{jjOPEs}
j^a_z(z) j^b_z(0) & = :j^a_z(z) j^b_z(0): + f^2 \eta^2 \frac{\kappa^{ab}}{z^2} 
+ f^2 \eta (2-\eta) \frac{i {f^{ab}}_c }{z} \frac{j^c_z(z)+j^c_z(0)}{2} \cr
& \qquad \qquad + f^2 \eta^2 \frac{i {f^{ab}}_c  \bar z}{z^2} \frac{j^c_{\bar z}(z)+j^c_{\bar z}(0)}{2}
+ ... \cr
j^a_{\bar z}(z) j^b_{\bar z}(0) & = :j^a_{\bar z}(z) j^b_{\bar z}(0): + f^2(1-\eta)^2\frac{\kappa^{ab}}{{\bar z}^2} 
+ f^2 (1-\eta)(1+\eta) \frac{i {f^{ab}}_c}{\bar z} \frac{ j^c_{\bar z}(z)+ j^c_{\bar z}(0)}{2} \cr
& \qquad \qquad 
+ f^2 (1-\eta)^2 \frac{i {f^{ab}}_c  z}{{\bar z}^2}\frac{j^c_{z}(z)+j^c_{z}(0)}{2}
+ ... \cr
j^a_z(z) j^b_{\bar z}(0) & = :j^a_z(z) j^b_{\bar z}(0): - f^2\eta(1-\eta)\kappa^{ab} 2 \pi \delta(z) 
+ f^2 (1-\eta)^2 \frac{i {f^{ab}}_c j^c_z(z)}{\bar z} \cr
& \qquad \qquad 
+ f^2 \eta^2 \frac{i {f^{ab}}_c j^c_{\bar z}(0)}{z}
+ ... 
\end{align}
%
The matrix $\kappa^{ab}$ is the invariant non-degenerate metric on the supergroup.
The ellipses contain subleading singular terms multiplying operators of dimension greater than one. 
The subleading terms involving composite operators built out of two currents were computed in \cite{Benichou:2010rk}. 
There it was also shown that all the singular terms in the current-current OPEs can be recursively deduced from \eqref{jjOPEs} by demanding consistency with current conservation and the Maurer-Cartan equation.
Notice that the currents in the first-order poles in \eqref{jjOPEs} are evaluated either at the point $z$ or $0$. The choice is guided%
\footnote{The computation of the terms involving one derivative of the currents performed in \cite{Ashok:2009xx} is not enough to fix completely the points at which the currents are evaluated. For instance some currents in \eqref{jjOPEs} could be evaluated at the point $\frac{z}{2}$. Here we make the choice that is arguably the most natural. A different choice would modify some formulas in the bulk of this paper, but the conclusions of section \ref{proofHirota} remain valid as long as the symmetry of the OPEs under the exchange of the operators on the left-hand side is preserved.} %
 by the study of the subleading terms involving derivatives of currents that was performed in \cite{Ashok:2009xx}. 

The holomorphic and anti-holomorphic components of the stress-tensor can be written in terms of the left-currents:
\be T = \frac{1}{2 f^2 \eta^2} \kappa_{ab}:j^a_z j^b_z: \qquad ; \qquad \bar T = \frac{1}{2 f^2 (1-\eta)^2} \kappa_{ab}:j^a_{\bar z} j^b_{\bar z}: \ee
The OPEs \eqref{jjOPEs} ensures that $T$ and $\bar T$ satisfy the canonical OPEs for the stress-tensor with central charge $c=\bar c$ equal to the superdimension of the supergroup, which is $-2$ for $PSl(n|n)$ \cite{Ashok:2009xx}\cite{Benichou:2010rk}.

\paragraph{Perturbative expansion.}
In section \ref{renormalization} and \ref{fusion} we will work in perturbation theory in the parameter $f^2$.
This corresponds to a semi-classical, large-radius expansion in the target space.
With our choice of normalization for the currents, all singular terms in the current-current OPEs are of order $f^2$ 
(including the sub-leading singular terms discussed in \cite{Benichou:2010rk} that are not explicitly written down in \eqref{jjOPEs}).
This makes the expansion in $f^2$ rather straightforward: computations at first order in $f^2$ involve one single OPE, computations at second order involve two OPEs, and so on.
Notice that the coefficients given in \eqref{jjOPEs} are exact to all orders in $f^2$ \cite{Ashok:2009xx}.


\subsection{Flat connections}

Let us consider the one-parameter family of connections:
\be A(\alpha;z) = \frac{2}{1+\alpha}  j^a_z(z) t_a dz +  \frac{2}{1-\alpha}  j^a_{\bar z}(z) t_a d\bar z \ee
with $\alpha$ a complex parameter.
Current conservation \eqref{CC} and the Maurer-Cartan equation \eqref{MC} ensure that this connection is flat:
\be dA(\alpha;z) + A(\alpha;z) \wedge A(\alpha;z) = 0\ee
As a consequence the following line operator:
\be T^{b, a}(\alpha) = P \exp\left(-\int_a^b dz A(\alpha;z) \right) \ee
does not depend on the integration path between the points $a$ and $b$%
\footnote{We use the same small-case latin characters $a,b,c...$ to denote both the endpoints of the path associated to transition matrices, and the super-algebra adjoint indices.}.
The symbol $P$ stands for path ordering.

For convenience we will only consider line operators defined on constant-time contours.
So from now on only the space component of the Lax connection will appear.
For simplicity we will also denote the space-component of the Lax connection by $A(\alpha;z)$.
In order to evaluate the relevant quantum effects, we will need the OPE between two connections.
The two connections can be taken in different representations $R$ and $R'$, and with different values of the spectral parameter.
The OPE follows from \eqref{jjOPEs}:
\begin{align} \label{OPEAA'} A_{R}(\alpha;z) A_{R'}(\beta;w) & = 
:A_{R}(\alpha;z) A_{R'}(\beta;w):  + a(\alpha,\beta;z-w) \cr 
& + b_{1,c}(\alpha,\beta;z-w) j^c_{z}(z) + b_{2,c}(\alpha,\beta;z-w) j^c_{z}(w) \cr
& + \bar b_{1,c}(\alpha,\beta;z-w) j^c_{\bar z}(z) + \bar b_{2,c}(\alpha,\beta;z-w) j^c_{\bar z}(w)
+... \end{align}
with:
\begin{align} & a(\alpha,\beta;z-w) = f^2\kappa^{ab}t^a_R t^b_{R'}\left\{  \frac{4}{(1+\alpha)(1+\beta)} \eta^2 \frac{1}{(z-w)^2}   
 + \frac{4}{(1-\alpha)(1-\beta)}  (1-\eta)^2 \frac{1}{(\bar z-\bar w)^2} \right. \cr 
& \qquad \left. - \left( \frac{4}{(1-\alpha)(1+\beta)}  + \frac{4}{(1+\alpha)(1-\beta)}   \right) \eta(1-\eta) 2 \pi \delta(z-w) \right\}  
\cr
& b_{1,c}(\alpha,\beta;z-w) = f^2 i {f^{ab}}_c t^a_R t^b_{R'} \left\{ \frac{1}{2}\frac{4}{(1+\alpha)(1+\beta)}  \eta(2-\eta) \frac{1 }{z-w}   
\right. \cr 
&  \qquad \left.
+\frac{1}{2}\frac{4}{(1-\alpha)(1-\beta)}   (1-\eta)^2 \frac{  (z-w)}{(\bar z-\bar w)^2}
+ \frac{4}{(1+\alpha)(1-\beta)} (1-\eta)^2 \frac{1 }{\bar z-\bar w} \right\}
\cr
& b_{2,c}(\alpha,\beta;z-w) = f^2 i {f^{ab}}_c t^a_R t^b_{R'} \left\{ \frac{1}{2}\frac{4}{(1+\alpha)(1+\beta)}  \eta(2-\eta) \frac{1 }{z-w}   
\right. \cr 
& \qquad \left.
+\frac{1}{2} \frac{4}{(1-\alpha)(1-\beta)}   (1-\eta)^2 \frac{  (z-w)}{(\bar z-\bar w)^2}
+ \frac{4}{(1-\alpha)(1+\beta)} (1-\eta)^2 \frac{1 }{\bar z-\bar w} \right\}
\cr
& \bar b_{1,c}(\alpha,\beta;z-w) =  f^2 i {f^{ab}}_c t^a_R t^b_{R'} \left\{  \frac{1}{2}\frac{4}{(1+\alpha)(1+\beta)}  \eta^2 \frac{(\bar z - \bar w) }{(z-w)^2}  \right. \cr 
& \qquad \left.
 + \frac{1}{2} \frac{4}{(1-\alpha)(1-\beta)}   (1-\eta)(1+\eta) \frac{1 }{\bar z-\bar w}
+ \frac{4}{(1-\alpha)(1+\beta)}  \eta^2 \frac{1 }{z-w}  \right\}
\cr
& \bar b_{2,c}(\alpha,\beta;z-w) =  f^2 i {f^{ab}}_c t^a_R t^b_{R'} \left\{  \frac{1}{2}\frac{4}{(1+\alpha)(1+\beta)}  \eta^2 \frac{(\bar z - \bar w) }{(z-w)^2}  \right. \cr 
& \qquad \left.
 + \frac{1}{2} \frac{4}{(1-\alpha)(1-\beta)}   (1-\eta)(1+\eta) \frac{1 }{\bar z-\bar w}
+ \frac{4}{(1+\alpha)(1-\beta)}  \eta^2 \frac{1 }{z-w}  \right\}
\end{align}
%


\subsection{Commutators of equal-time connections and $r,s$ matrices}\label{rsMatrices}

From the OPE \eqref{OPEAA'} we deduce the commutator between two equal-time connections (see Appendix \ref{OPE[]}):
\begin{align}\label{[A,A']v1} & [ A_{R}(\alpha;\sigma_1), A_{R'}(\beta;\sigma_2) ] 
\cr & \quad
= 2\pi i f^2 t_{a,R} t_{b,R'} \left\{
 \kappa^{ab}   \left(  \frac{4}{(1+\alpha)(1+\beta)} \eta^2 - \frac{4}{(1-\alpha)(1-\beta)} (1-\eta)^2 \right) \delta'(\sigma_1-\sigma_2) \right. \cr
&\qquad +i {f^{ab}}_c j_z^c(\sigma_2) \left( - \frac{4}{(1+\alpha)(1+\beta)} + (1-\eta)^2 \frac{16}{(1+\alpha)(1+\beta)(1-\alpha)(1-\beta)} \right) \delta(\sigma_1-\sigma_2) \cr
&\qquad  \left. + i {f^{ab}}_c j_{\bar z}^c(\sigma_2) \left( \frac{4}{(1-\alpha)(1-\beta)} - \eta^2 \frac{16}{(1+\alpha)(1+\beta)(1-\alpha)(1-\beta)} \right) \delta(\sigma_1-\sigma_2) \right \}
 \end{align}
%
Notice that the sub-leading singular terms in the current-current OPEs, that are contained in the ellipses in \eqref{jjOPEs}, do not contribute to this commutator.
Indeed locality imposes that equal-time operators commute if they are sitting at different positions.
Hence all the terms on the right-hand side of the commutator come with a delta-function, or a derivative thereof. 
All sub-leading singular terms in the current-current OPEs involve operators of dimension strictly greater than one. So dimensional analysis forbids the appearance of these operators in the commutator. This point is further discussed at the end of appendix \ref{Maillet}.

The coefficients of the current-current OPEs given in \eqref{jjOPEs} have been computed from first principle to all orders in $f^2$ in \cite{Ashok:2009xx}. This implies that the commutator \eqref{[A,A']v1} is exact to all orders in $f^2$.

Using:
\be i {f^{ab}}_c  t_{a,R} t_{b,R'} = [t_{c,R'}, t_{a,R} t_{b,R'}\kappa^{ab}] = - [t_{c,R}, t_{a,R} t_{b,R'}\kappa^{ab}] \ee
we can rewrite the commutator as:
\begin{align}\label{[A,A']} [ A_{R}(\alpha;\sigma_1), A_{R'}(\beta;\sigma_2) ]
&= 2\pi i f^2 Q \left(  \frac{4}{(1+\alpha)(1+\beta)} \eta^2 - \frac{4}{(1-\alpha)(1-\beta)} (1-\eta)^2 \right) \delta'(\sigma_1-\sigma_2) \cr
& \quad + 2\pi i f^2  \left[ N_R A_{R}(\alpha;\sigma_1) +  N_{R'} A_{R'}(\beta;\sigma_2) ,Q \right] \delta(\sigma_1-\sigma_2)\end{align}
where we introduced the matrix $Q$ defined as:
\be Q = \kappa^{ab} t_{a,R} t_{b,R'} \ee
and $N_R$ and $N_{R'}$ are solutions of:
\begin{align}
& -\frac{2}{1+\alpha}N_R + \frac{2}{1+\beta}N_{R'} =  - \frac{4}{(1+\alpha)(1+\beta)} + (1-\eta)^2 \frac{16}{(1+\alpha)(1+\beta)(1-\alpha)(1-\beta)}
\cr
& -\frac{2}{1-\alpha}N_R + \frac{2}{1-\beta}N_{R'} = \frac{4}{(1-\alpha)(1-\beta)} - \eta^2 \frac{16}{(1+\alpha)(1+\beta)(1-\alpha)(1-\beta)}
\end{align}
The determinant of the system is zero when $\alpha=\beta$.
For $\alpha \neq \beta$ we obtain:
\be\label{NN'} N_R = \frac{2}{\alpha-\beta} \frac{(1+\beta-2\eta)^2}{(1+\beta)(1-\beta)} \qquad ; \qquad
N_{R'} = \frac{2}{\alpha-\beta} \frac{(1+\alpha-2\eta)^2}{(1+\alpha)(1-\alpha)}
 \ee
We notice that the commutator \eqref{[A,A']} matches a $(r,s)$ Maillet system (cf equation \eqref{PoissonJJ}):
\begin{align} [ A_{R}(\alpha;\sigma_1), A_{R'}(\beta;\sigma_2) ] & = 2s \delta'(\sigma_1-\sigma_2) \cr
&  + \left[ A_{R}(\alpha;\sigma_1) + A_{R'}(\beta;\sigma_2), r \right] \delta(\sigma_1-\sigma_2) \cr
&  + \left[ A_{R}(\alpha;\sigma_1) - A_{R'}(\beta;\sigma_2), s \right] \delta(\sigma_1-\sigma_2)
 \end{align}
with:
\begin{align} r = 2\pi i f^2 \frac{N_{R}+N_R'}{2} Q \qquad ; \qquad
s = 2\pi i f^2 \frac{N_{R}-N_R'}{2} Q 
\end{align}
Explicitly the matrices $r$ and $s$ read:
\begin{align}\label{rsMat}
r & = \pi i f^2  \frac{2}{\alpha-\beta} \left[   \frac{(1+\beta-2\eta)^2}{(1+\beta)(1-\beta)} + \frac{(1+\alpha-2\eta)^2}{(1+\alpha)(1-\alpha)}\right]Q \cr
s & = \pi i f^2  \left[\frac{4}{(1+\alpha)(1+\beta)} \eta^2 - \frac{4}{(1-\alpha)(1-\beta)} (1-\eta)^2 \right]Q \end{align}
The matrices $r\pm s$ satisfy the extended classical Yang-Baxter equation \cite{Maillet:1985fn} for all values of the parameter $\eta$:
\be [(r\pm s)_{\alpha,\beta},(r\pm s)_{\alpha,\gamma}] + [(r\pm s)_{\alpha,\beta},(r\pm s)_{\beta,\gamma}] + [(r\pm s)_{\gamma,\beta},(r\pm s)_{\alpha,\gamma}] = 0 \ee
The system we are considering is of the type studied in \cite{Freidel:1991jv}.
Eventually notice that the matrix $Q$ that appears both in the $r$ and $s$ matrices can be written as:
\be Q = \frac{c^{(2)}_{R \otimes R'}-c^{(2)}_{R}-c^{(2)}_{R'}}{2} \ee
where $c^{(2)}_{R}$ is the quadratic Casimir evaluated in the representation $R$. The same matrix appeared in the computation of the conformal dimension of composite operators in \cite{Benichou:2010rk}.


\subsection{Transition, Monodromy and Transfer matrices}

Since the notations and the vocabulary are not homogeneous in the literature, let us pause to define the line operators that we will study.
The transition matrix between the points $a$ and $b$ in the representation $R$ is defined as follows:
\be T^{b,a}_R(\alpha) = P \exp\left(-\int_a^b A_R(\alpha) \right) \ee
Flatness of the connection implies that the transition matrix does not depend on the integration path chosen.
For stringy purposes we are led to consider the theory on a cylinder.
So we compactify the worldsheet space direction: $\sigma=\sigma+2\pi$.
The monodromy matrix is the transition matrix associated to a closed contour $(C)$ winding once around the cylinder:
\be \Omega_{R}^{(C)}(\alpha) = P \exp \left(-\oint_C A_R(\alpha) \right) \ee
Eventually the transfer matrix is the trace of the monodromy matrix:
\be \T_{R}^{(C)}(\alpha) = STr\ \Omega_{R}^{(C)}(\alpha) \ee
%


\section{Divergences in line operators}\label{renormalization}

In a quantum theory the line operators are generically ill-defined because of the singularities encountered when the integrated connections collide.
To properly define a transition matrix we have to regularize and renormalize it.
This is the problem we consider in this section.
To compute the UV divergences appearing in a transition matrix it is convenient to expand it as:
\be\label{expansionT} T^{b,a}(\alpha) = \sum_{N=0}^{\infty} (-1)^N T^{b,a}_N(\alpha) \ee
with:
\be\label{T_N} T^{b,a}_N(\alpha) = \frac{1}{N!} P \left( \int_a^b A(\alpha) \right)^N 
= \int_{b>\sigma_1>...>\sigma_N>a}d\sigma_1...d\sigma_N A^{a_1}(\alpha;\sigma_1)...A^{a_N}(\alpha;\sigma_N) t_{a_1}...t_{a_N} \ee

\subsection{Regularization}\label{regularization}

To regularize the UV divergences that appear when two connections collide we choose a ``principal-value'' regularization scheme as suggested in \cite{Mikhailov:2007eg}.
The singularities encountered in the OPE of two equal-time connections are regularized by an infinitesimal shift one of the connections in the time direction, in a symmetric way:
\be A_{R}^\alpha(\sigma_1) A_{R'}^\beta(\sigma_2) \longrightarrow
 \frac{1}{2} \left(A_{R}^\alpha(\sigma_1+i \epsilon) A_{R'}^\beta(\sigma_2) + A_{R}^\alpha(\sigma_1-i \epsilon) A_{R'}^\beta(\sigma_2) \right) \ee
where $\epsilon$ is the UV regulator.
Explicitly the singularities encountered in the OPEs become:
\be  \frac{1}{\sigma} \longrightarrow P.V. \frac{1}{\sigma} =  \frac{1}{2}\left(\frac{1}{\sigma+i\epsilon}+\frac{1}{\sigma-i\epsilon}\right) = \frac{\sigma}{\sigma^2+\epsilon^2} = \p_\sigma \frac{1}{2} \log(\sigma^2+\epsilon^2) \ee
\be \frac{1}{\sigma^2 } \longrightarrow P.V. \frac{1}{\sigma^2} =  \frac{1}{2}\left(\frac{1}{(\sigma+i\epsilon)^2}+\frac{1}{(\sigma-i\epsilon)^2}\right) = \frac{\sigma^2-\epsilon^2}{(\sigma^2+\epsilon^2)^2} = - \p_\sigma P.V. \frac{1}{\sigma} \ee
This regularization scheme appears naturally when considering the fusion of transition matrices, as will be explained in section \ref{fusion}.
Different prescriptions to regularize the transition matrices have been considered in the literature. In \cite{Mikhailov:2007mr} a sharp regularization was used: the distance between the integrated operators was constrained to be greater than a minimal length. In \cite{Bachas:2004sy}\cite{Bachas:2009mc} a smooth regularization was performed at the level of the Fourier modes of the connection. 
In appendix \ref{divWZW} we rederive some of the results of \cite{Bachas:2004sy} using the regularization scheme presented above.
In appendix \ref{altRegScheme} we study the divergences appearing in transfer matrices using the regularization scheme of \cite{Mikhailov:2007mr}.


\subsection{Divergences at first-order in perturbation theory}\label{divOrder1}

First we study the divergences that appear at first order in perturbation theory.
Accordingly we consider a single OPE between two connections.
We write the regularized OPE between two equal-time connections as:
\begin{align}\label{OPEAAreg} A^a(\alpha;\sigma) A^b(\alpha;\sigma') & = p_2 \kappa^{ab} P.V. \frac{1}{(\sigma-\sigma')^2} 
 + p_1 {f^{ab}}_c (j_z^c(\sigma)+j_z^c(\sigma')) P.V.\frac{1}{\sigma-\sigma'} \cr
 & + \bar p_1 {f^{ab}}_c (j_{\bar z}^c(\sigma)+j_{\bar z}^c(\sigma')) P.V.\frac{1}{\sigma-\sigma'} + ... \end{align}
The numerical coefficients $p_2$, $p_1$, $\bar p_1$ can be read directly from the OPE \eqref{OPEAA'}. Their explicit value is not needed in the following.

\begin{figure}
\centering
\includegraphics[scale=0.55]{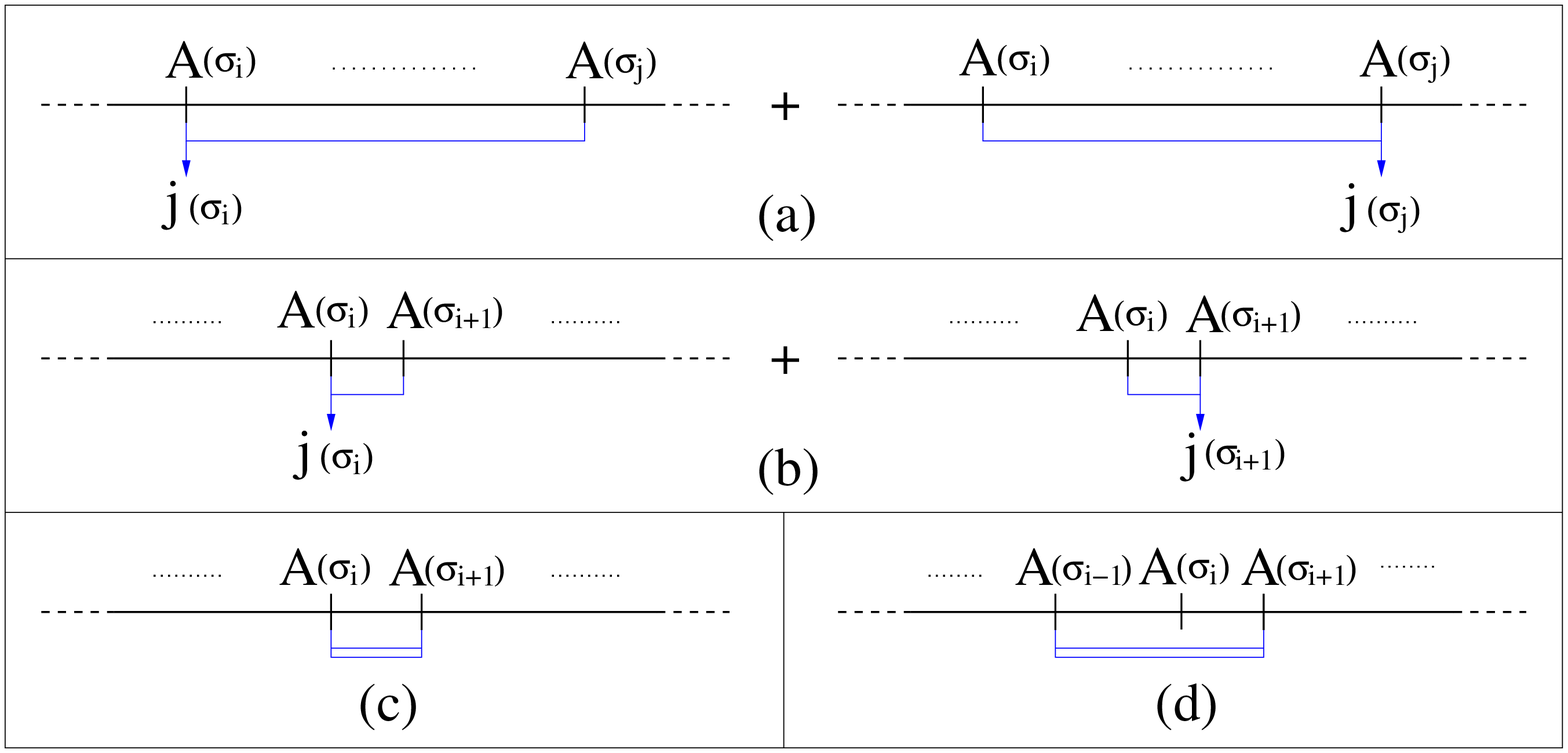}
\caption{OPEs potentially producing divergences at first order. The blue line represents the OPE under consideration. An arrow indicates the point at which the resulting currents are evaluated. A double line means that we consider the second-order pole in the OPE. \label{div1}}
\end{figure}

\paragraph{First-order poles.}
First let us consider the divergences coming from the first-order singularities in \eqref{OPEAAreg}.
We perform the OPE between two connections $A^{a_i}(\alpha;\sigma_i)$ and $A^{a_j}(\alpha;\sigma_j)$ in the operator $T^{b,a}_N(\alpha)$ defined in \eqref{T_N}, and isolate the contribution of the first-order poles (see Figure \ref{div1}(a)). We obtain:
\begin{align}
& \int_{b>\sigma_1>...>\sigma_N>a}d\sigma_1...d\sigma_N 
A^{a_1}(\alpha;\sigma_1)...A^{a_{i-1}}(\alpha;\sigma_{i-1})A^{a_{i+1}}(\alpha;\sigma_{i+1})...A^{a_{j-1}}(\alpha;\sigma_{j-1})\cr
& \quad \times A^{a_{j+1}}(\alpha;\sigma_{j+1})...A^{a_N}(\alpha;\sigma_N) 
(p_1 j_z^{c}(\sigma_i) + p_1 j_z^c(\sigma_j) + \bar p_1 j_{\bar z}^c(\sigma_i) + \bar p_1 j_{\bar z}^c(\sigma_j)) \cr
& \quad \times \left(P.V.\frac{1}{\sigma_i-\sigma_j}\right) {f^{a_i a_j}}_c
t_{a_1}...t_{a_N}
\end{align}
The integral over the free coordinate is easily evaluated, for instance:
\be \int_{\sigma_{i+1}}^{\sigma_{i-1}} d\sigma_i P.V. \frac{1}{\sigma_i-\sigma_j} 
= \frac{1}{2} \log \left( \frac{(\sigma_j-\sigma_{i-1})^2 + \epsilon^2}{(\sigma_j-\sigma_{i+1})^2 + \epsilon^2} \right) \ee
Thus we obtain a logarithmic divergence if and only if the two connections are adjacent: $j=i\pm 1$.
This is the case depicted in Figure \ref{div1}(b).
But in that case the generators $t_{a_i}$ and $t_{a_j}$ in $T_N^{b,a}(\alpha)$ are also adjacent,
and the contraction of a structure constant with the product of two generators vanishes:
\be {f^{ab}}_c t_a t_b = \frac{i}{2} {f^{ab}}_c {f_{ab}}^d t_d = 0 \ee
We deduce that there is no divergence coming from the first-order poles in \eqref{OPEAAreg}.

\paragraph{Second-order poles.}
Now let us consider the divergences coming from the second-order pole in \eqref{OPEAAreg}. 
First we consider the OPE between two adjacent connections $A^{a_i}(\alpha;\sigma_i)$ and $A^{a_{i+1}}(\alpha;\sigma_{i+1})$ in $T^{b,a}_N(\alpha)$, and isolate the contribution of the second-order pole (see Figure \ref{div1}(c)):
\begin{align}
& \int_{b>\sigma_1>...>\sigma_N>a}d\sigma_1...d\sigma_N 
A^{a_1}(\alpha;\sigma_1)...A^{a_{i-1}}(\alpha;\sigma_{i-1})A^{a_{i+2}}(\alpha;\sigma_{i+2})...A^{a_N}(\alpha;\sigma_N)  \cr
& \quad \times p_2 \kappa^{a_i a_{i+1}} P.V.\frac{1}{(\sigma_i-\sigma_{i+1})^2} 
t_{a_1}...t_{a_N}
\end{align}
We compute the double integral over the free coordinates to obtain:
\begin{align} & \int_{\sigma_{i+2}}^{\sigma_{i-1}} d\sigma_{i+1} \int_{\sigma_{i+1}}^{\sigma_{i-1}} d\sigma_i P.V. \frac{1}{(\sigma_i-\sigma_{i+1})^2}
= \int_{\sigma_{i+2}}^{\sigma_{i-1}} d\sigma_{i+1} P.V. \frac{1}{\sigma_{i+1}-\sigma_{i-1}} + 0
 =  \log \epsilon + \mathrm{finite}
\end{align}
The linear divergences cancel in our regularization scheme.
Notice that the contraction of the metric with the product of two generators gives the quadratic Casimir operator $C^{(2)}$, which commute with all generators. Hence these logarithmic divergences we get from $T^{b,a}_N(\alpha)$ add up to:
\be\label{divTo1t1} (N-1) p_2 \log \epsilon C^{(2)} T^{b,a}_{N-2}(\alpha) \ee
The factor of $N-1$ comes from the $N-1$ different pairs of adjacent connections in the operator $T^{b,a}_N(\alpha)$.

Next we consider the OPE of two connections $A^{a_{i-1}}(\alpha;\sigma_{i-1})$ and $A^{a_{i+1}}(\alpha;\sigma_{i+1})$ separated by a third one, and isolate the contribution of the double pole (see Figure \ref{div1}(d)). We obtain:
\begin{align}
& \int_{b>\sigma_1>...>\sigma_N>a}d\sigma_1...d\sigma_N 
A^{a_1}(\alpha;\sigma_1)...A^{a_{i-2}}(\alpha;\sigma_{i-2})A^{a_{i}}(\alpha;\sigma_{i})A^{a_{i+2}}(\alpha;\sigma_{i+2})...A^{a_N}(\alpha;\sigma_N)  \cr
& \quad \times p_2 \kappa^{a_{i-1} a_{i+1}} P.V.\frac{1}{(\sigma_{i-1}-\sigma_{i+1})^2} 
t_{a_1}...t_{a_N}
\end{align}
We compute the double integral over the free coordinates to obtain:
\begin{align} & \int_{\sigma_{i+2}}^{\sigma_{i}} d\sigma_{i+1} \int_{\sigma_{i}}^{\sigma_{i-2}} d\sigma_{i-1} P.V. \frac{1}{(\sigma_{i-1}-\sigma_{i+1})^2}
 = - \log \epsilon + \mathrm{finite}
\end{align}
The contraction of the metric with the generators also simplifies:
\be \kappa^{a_{i-1} a_{i+1}} t_{a_{i-1}} t_{a_i} t_{a_{i+1}} = C^{(2)} t_{a_i} + \kappa^{a_{i-1} a_{i+1}} i {f_{a_i a_{i+1}}}^b t_{a_{i-1}} t_b = C^{(2)} t_{a_i} \ee
Hence these logarithmic divergences we obtained from $T^{b,a}_N(\alpha)$ add up to:
\be\label{divTo1t2} - (N-2) p_2 \log \epsilon C^{(2)} T^{b,a}_{N-2}(\alpha) \ee
The factor of $N-2$ comes from the $N-2$ different pairs of connections separated by a third one in the operator $T^{b,a}_N(\alpha)$.
OPEs between connections separated by more than one other connections do not produce any divergences.

\paragraph{Upshot.}
Summing the divergences obtained in \eqref{divTo1t1} and \eqref{divTo1t2} we obtain that first-order quantum effects lead to a logarithmic divergence in the transition matrix of the form:
\be\label{logDivO1} p_2 C^{(2)} \log \epsilon T^{b,a}(\alpha)  \ee
%


\subsection{Divergences at second-order in perturbation theory}\label{divOrder2}

Now we compute the divergences appearing in the transition matrix at second order in perturbation theory.
So we perform two OPEs. These OPEs can be taken between two distinct pairs of connections. But we can also perform a first OPE between two connections, and then take the OPE of the resulting currents with a third connection. 
We will evaluate these two types of terms in turn.
For the latter case, that we loosely call a triple OPE, we will need the OPE between a current and a connection:
\begin{align}\label{OPEjAreg} j_z^a(\sigma) A^a(\alpha;\sigma') & = q_2 \kappa^{ab} P.V. \frac{1}{(\sigma-\sigma')^2} 
 + q_1 {f^{ab}}_c (j_z^c(\sigma)+j_z^c(\sigma')) P.V.\frac{1}{\sigma-\sigma'} \cr
 & + \bar q_1 {f^{ab}}_c (j_{\bar z}^c(\sigma)+j_{\bar z}^c(\sigma')) P.V.\frac{1}{\sigma-\sigma'} + ... \end{align}
\begin{align}\label{OPEjbAreg} j_{\bar z}^a(\sigma) A^a(\alpha;\sigma') & = r_2 \kappa^{ab} P.V. \frac{1}{(\sigma-\sigma')^2} 
 + r_1 {f^{ab}}_c (j_z^c(\sigma)+j_z^c(\sigma')) P.V.\frac{1}{\sigma-\sigma'} \cr
 & + \bar r_1 {f^{ab}}_c (j_{\bar z}^c(\sigma)+j_{\bar z}^c(\sigma')) P.V.\frac{1}{\sigma-\sigma'} + ... \end{align}
The coefficients can be easily deduced from the current algebra \eqref{jjOPEs}.
Their explicit value will not be needed in the following.

\paragraph{OPEs between distinct pairs of connections.}
First let us perform two OPEs between two distinct pairs of connections. From the analysis of section \ref{divOrder1} the result is straightforward. 
We obtain a logarithmic divergence equal to:
\be\label{logDivO2} \frac{1}{2}(p_2 C^{(2)} \log \epsilon)^2 T^{b,a}(\alpha) \ee
The factor of $\frac{1}{2}$ comes from the fact that the order in which we perform the two OPEs does not matter.

\paragraph{Triple OPE: contribution of the first-order pole.}
Now we consider triple OPEs. We first perform an OPE between two connexions $A^{a_i}(\alpha;\sigma_i)$ and $A^{a_j}(\alpha;\sigma_j)$ in the operator $T^{b,a}_N(\alpha)$. These two connections need to be separated by at least one other connection, else the result vanishes already.
Then we perform a second OPE between the resulting currents and a third connection. 

\begin{figure}
\centering
\includegraphics[scale=0.55]{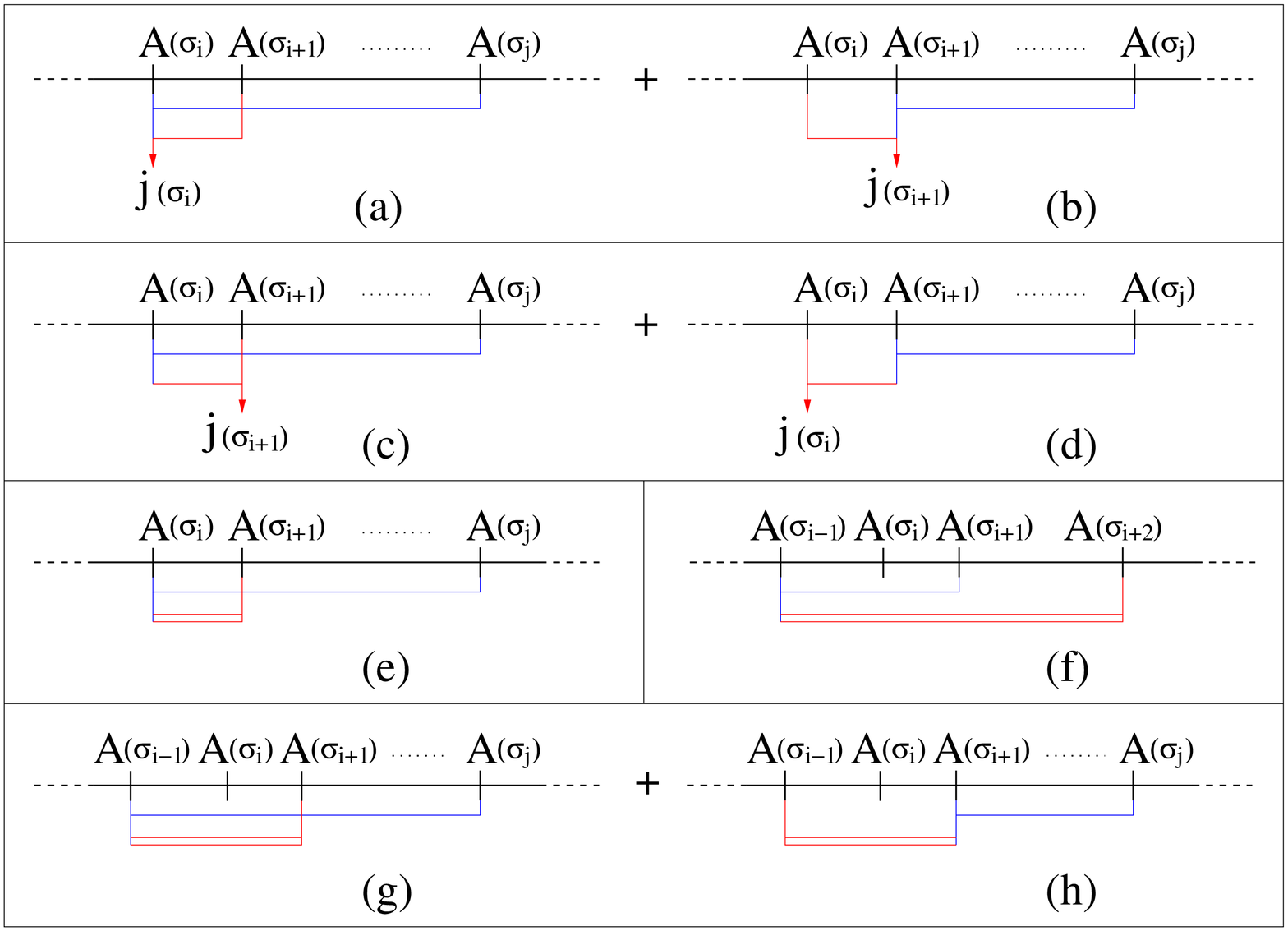}
\caption{Triple OPEs potentially producing divergences at second order. The blue line represents the first OPE, and the red line represents the second OPE. An arrow indicates the point at which the currents are evaluated. A double line means that we consider the second-oder pole in the OPE. \label{div2}}
\end{figure}

Let us evaluate the contribution of the first-order poles in the second OPE. 
The intermediate currents are evaluated either at $\sigma_i$ or at $\sigma_j$. For a divergence to appear, we have to take the OPE between the currents and one of the neighboring connections. Let us take a definite example: we consider the intermediate currents evaluated at $\sigma_i$. Then we perform the OPE of these currents with the connection $A^{a_{i+1}}(\alpha;\sigma_{i+1})$, and isolate the contribution of the first-order pole in this OPE that comes with a current evaluated at $\sigma_i$ (see Figure \ref{div2}(a)). We obtain:
\begin{align}
& \int_{b>\sigma_1>...>\sigma_N>a}d\sigma_1...d\sigma_N 
A^{a_1}(\alpha;\sigma_1)...A^{a_{i-1}}(\alpha;\sigma_{i-1})A^{a_{i+2}}(\alpha;\sigma_{i+2})...A^{a_{j-1}}(\alpha;\sigma_{j-1})\cr
& \quad \times A^{a_{j+1}}(\alpha;\sigma_{j+1})...A^{a_N}(\alpha;\sigma_N) 
((p_1 q_1 + \bar p_1 r_1)j_z^{d}(\sigma_i) + (p_1 \bar q_1 + \bar p_1 \bar r_1) j_{\bar z}^d(\sigma_i) ) \cr
& \quad \times P.V.\frac{1}{\sigma_i-\sigma_j} P.V.\frac{1}{\sigma_i-\sigma_{i+1}} {f^{a_i a_j}}_c {f^{c a_{i+1}}}_d
t_{a_1}...t_{a_N}
\end{align}
We perform the integral over $\sigma_{i+1}$:
\be \int_{\sigma_{i+2}}^{\sigma_i}d\sigma_{i+1}  P.V.\frac{1}{\sigma_i-\sigma_{i+1}} = -\log \epsilon + \mathrm{finite} \ee
We can combine the previous term with the one obtained in the following way: the first OPE is taken between $A^{a_{i+1}}(\alpha;\sigma_{i+1})$ and $A^{a_j}(\alpha;\sigma_j)$, and we consider only the currents evaluated at $\sigma_{i+1}$. Then we perform the OPE between these currents and the connection $A^{a_i}(\alpha;\sigma_i)$, and isolate the part proportional to the currents evaluated at $\sigma_{i+1}$ (see Figure \ref{div2}(b)). We obtain:
\begin{align}
& \int_{b>\sigma_1>...>\sigma_N>a}d\sigma_1...d\sigma_N 
A^{a_1}(\alpha;\sigma_1)...A^{a_{i-1}}(\alpha;\sigma_{i-1})A^{a_{i+2}}(\alpha;\sigma_{i+2})...A^{a_{j-1}}(\alpha;\sigma_{j-1})\cr
& \quad \times A^{a_{j+1}}(\alpha;\sigma_{j+1})...A^{a_N}(\alpha;\sigma_N) 
((p_1 q_1 + \bar p_1 r_1)j_z^{d}(\sigma_i) + (p_1 \bar q_1 + \bar p_1 \bar r_1) j_{\bar z}^d(\sigma_i) ) \cr
& \quad \times P.V.\frac{1}{\sigma_{i+1}-\sigma_j} P.V.\frac{1}{\sigma_{i+1}-\sigma_{i}} {f^{a_{i+1} a_j}}_c {f^{c a_{i}}}_d
t_{a_1}...t_{a_N}
\end{align}
We perform the integral over $\sigma_i$:
\be \int_{\sigma_{i+1}}^{\sigma_{i-1}} d\sigma_{i} P.V.\frac{1}{\sigma_{i+1}-\sigma_{i}} = +\log \epsilon + \mathrm{finite} \ee
Then we add these two terms. Thanks to the Jacobi identity, they simplify:
\be -{f^{a_i a_j}}_c {f^{c a_{i+1}}}_d +  {f^{a_{i+1} a_j}}_c {f^{c a_{i}}}_d = {f^{a_{i+1} a_i}}_c {f^{c a_{j}}}_d \ee
But we encounter once again the contraction between a structure constant and the product of two generators, which vanishes.

Similarly the divergent terms obtained from the OPEs depicted in Figure \ref{div2}(c) and \ref{div2}(d) cancel when combined together.
Thus all terms obtained in this way cancel by pairs. These terms produce no new divergence in the transition matrix.

\paragraph{Triple OPE: contribution of the second-order pole.}
Now we consider the contribution of the second-order pole in the OPE between the intermediate currents and the third connection.
First let us consider the case where the intermediate current is adjacent to the third connection. For definiteness we consider the following case: the first OPE is taken between the two connexions $A^{a_i}(\alpha;\sigma_i)$ and $A^{a_j}(\alpha;\sigma_j)$ in the operator $T^{b,a}_N(\alpha)$. We isolate the intermediate current evaluated at the point $\sigma_i$. Then we perform the OPE between these currents and the connection $A^{a_{i+1}}(\alpha;\sigma_{i+1})$ and isolate the contribution of the double pole in this OPE (see Figure \ref{div2}(e)). This term is proportional to:
\be {f^{a_i a_j}}_d \kappa^{d a_{i+1}} = f^{a_i a_j a_{i+1}} \ee
Once contracted with the generators, this tensor gives zero.

Next we consider the case where the intermediate currents and the third connection are separated by one other connection. The first OPE is taken between the two connexions $A^{a_{i-1}}(\alpha;\sigma_{i-1})$ and $A^{a_j}(\alpha;\sigma_j)$ in the operator $T^{b,a}_N(\alpha)$. We isolate the intermediate currents evaluated at the point $\sigma_{i-1}$. Then we perform the OPE between these currents and the connection $A^{a_{i+1}}(\alpha;\sigma_{i+1})$ and isolate the contribution of the double pole in this OPE (see Figure \ref{div2}(g)).
The result is:
\begin{align}
& \int_{b>\sigma_1>...>\sigma_N>a}d\sigma_1...d\sigma_N 
A^{a_1}(\alpha;\sigma_1)...A^{a_{i-2}}(\alpha;\sigma_{i-2})A^{a_{i}}(\alpha;\sigma_{i})A^{a_{i+2}}(\alpha;\sigma_{i+2})...A^{a_{j-1}}(\alpha;\sigma_{j-1})\cr
& \quad \times A^{a_{j+1}}(\alpha;\sigma_{j+1})...A^{a_N}(\alpha;\sigma_N)  
\cr & \quad \times 
(p_1 q_2+\bar p_1 r_2) P.V.\frac{1}{\sigma_{i-1}-\sigma_j} P.V.\frac{1}{(\sigma_{i+1}-\sigma_{i-1})^2} {f^{a_{i-1} a_j}}_c \kappa^{c a_{i+1}}
t_{a_1}...t_{a_N}
\end{align}
We perform the integration over $\sigma_{i-1}$ and $\sigma_{i+1}$:
\be \int_{\sigma_{i}}^{\sigma_{i-2}} d\sigma_{i-1} \int_{\sigma_{i+2}}^{\sigma_{i}} d\sigma_{i+1} P.V.\frac{1}{\sigma_{i-1}-\sigma_j} P.V.\frac{1}{(\sigma_{i+1}-\sigma_{i-1})^2}
= - \frac{\log \epsilon}{\sigma_i -\sigma_j} + \mathrm{finite} \ee
We want to combine this term this the following one: the first OPE is taken between the two connexions $A^{a_{i+1}}(\alpha;\sigma_{i+1})$ and $A^{a_j}(\alpha;\sigma_j)$ in the operator $T^{b,a}_N(\alpha)$. We isolate the intermediate currents evaluated at the point $\sigma_{i+1}$. Then we perform the OPE between these currents and the connection $A^{a_{i-1}}(\alpha;\sigma_{i-1})$ and isolate the contribution of the double pole in this OPE (see Figure \ref{div2}(h)).
The result is:
\begin{align}
& \int_{b>\sigma_1>...>\sigma_N>a}d\sigma_1...d\sigma_N 
A^{a_1}(\alpha;\sigma_1)...A^{a_{i-2}}(\alpha;\sigma_{i-2})A^{a_{i}}(\alpha;\sigma_{i})A^{a_{i+2}}(\alpha;\sigma_{i+2})...A^{a_{j-1}}(\alpha;\sigma_{j-1})\cr
& \quad \times A^{a_{j+1}}(\alpha;\sigma_{j+1})...A^{a_N}(\alpha;\sigma_N)  \cr
& \quad \times (p_1 q_2+\bar p_1 r_2) P.V.\frac{1}{\sigma_{i+1}-\sigma_j} P.V.\frac{1}{(\sigma_{i+1}-\sigma_{i-1})^2} {f^{a_{i+1} a_j}}_c \kappa^{c a_{i-1}}
t_{a_1}...t_{a_N}
\end{align}
We perform the integration over $\sigma_{i-1}$ and $\sigma_{i+1}$:
\be \int_{\sigma_{i}}^{\sigma_{i-2}} d\sigma_{i-1} \int_{\sigma_{i+2}}^{\sigma_{i}} d\sigma_{i+1}  P.V.\frac{1}{\sigma_{i+1}-\sigma_j} P.V.\frac{1}{(\sigma_{i+1}-\sigma_{i-1})^2}
= - \frac{\log \epsilon}{\sigma_i -\sigma_j} + \mathrm{finite} \ee
The divergent parts cancel in the sum of these two terms since:
\be {f^{a_{i+1} a_j}}_c \kappa^{c a_{i-1}} + {f^{a_{i-1} a_j}}_c \kappa^{c a_{i+1}} = 0 \ee
So these terms do not lead to any new divergence either.

There is one last configuration that may produce divergences.
The first OPE is taken between the two connexions $A^{a_{i-1}}(\alpha;\sigma_{i-1})$ and $A^{a_{i+1}}(\alpha;\sigma_{i+1})$ in the operator $T^{b,a}_N(\alpha)$. We isolate the intermediate current evaluated at the point $\sigma_{i-1}$. Then we perform the OPE between these currents and the connection $A^{a_{i+2}}(\alpha;\sigma_{i+2})$ and isolate the contribution of the double pole in this OPE (see Figure \ref{div2}(f)).
But this term is proportional to:
\be f^{a_{i-1}a_{i+1}a_{i+2}} \ee
which vanishes once contracted with the generators.

\paragraph{Renormalization of the transition matrices.}
At second order in perturbation theory the divergences appearing in the transition matrix are given in \eqref{logDivO2}.
The divergences coming from triple collisions add up to zero.
So up to this order all divergences can be canceled by a simple scalar wave-function renormalization of the transition matrix:
\be T^{b,a}(\alpha) \to T^{b,a}_{ren.}(\alpha) = e^{-p_2 C^{(2)} \log \epsilon} T^{b,a}(\alpha) + \mathcal{O}(f^6)\ee
In particular the cancellation of the divergences does not require a renormalization of the spectral parameter.


\subsection{The quantum Monodromy and Transfer matrices}

Now we consider the theory on a cylinder, and we study the divergences appearing in the monodromy matrix.
It is convenient to expand the monodromy matrix as:
\be \Omega(\alpha) = \sum_{N=0}^{\infty} (-1)^N \Omega_N(\alpha) \ee
with:
\be\label{M_N} \Omega_N(\alpha) = \frac{1}{N!} P \left( \oint A(\alpha) \right) 
= \int_{2\pi>\sigma_1>...>\sigma_N>0}d\sigma_1...d\sigma_N A^{a_1}(\alpha;\sigma_1)...A^{a_N}(\alpha;\sigma_N) t_{a_1}...t_{a_N} \ee
Notice that the monodromy matrix depends on the starting point of the integration path. Changing the starting point is equivalent to performing a similarity transformation on the monodromy matrix. For definiteness we consider a path that extends between $\sigma=0$ and $\sigma=2\pi$.

New divergences appear in the monodromy matrix with respect to the transition matrix studied previously. 
Indeed a connection sitting near the starting point of the integration path can now collide with another connection sitting near the endpoint of the integration path.
These are the potential sources of divergences that we study now. We start with the new divergences that appear at first-order.

\begin{figure}
\centering
\includegraphics[scale=0.55]{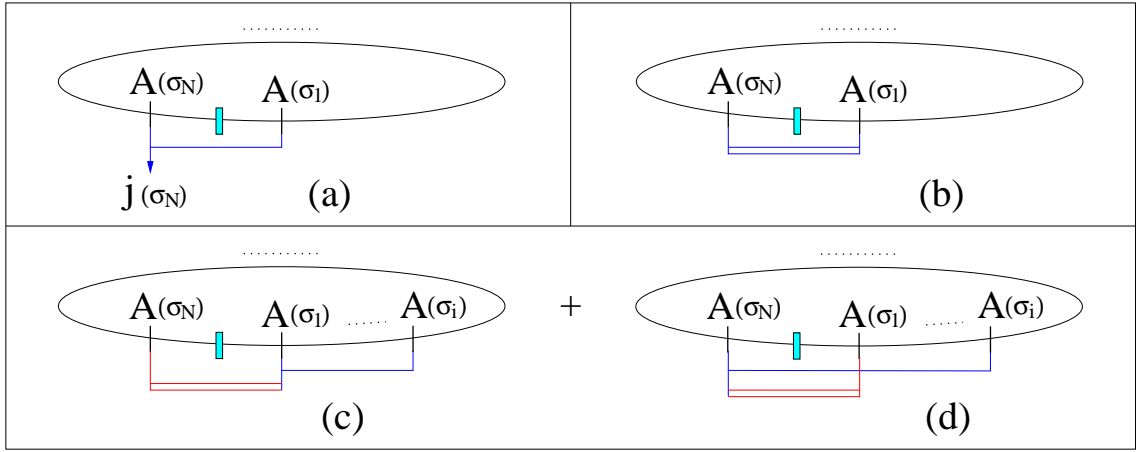}
\caption{OPEs potentially producing new divergences in the monodromy matrix. The blue line represents the first OPE, and the red line represents the second OPE. A double line means that we consider the second-order pole in the OPE. The light-blue rectangle represents the starting point of the integration path.\label{div2M}}
\end{figure}

\paragraph{First-order poles.}
Let us consider the operator $\Omega_N$ defined in \eqref{M_N}. We perform an OPE between $A^{a_1}(\alpha;\sigma_1)$ and $A^{a_N}(\alpha;\sigma_N)$, and isolate the contribution from the first-order poles (see Figure \ref{div2M}(a)). We obtain:
\begin{align}
& \int_{2\pi>\sigma_1>...>\sigma_N>0}d\sigma_1...d\sigma_N A^{a_2}(\alpha;\sigma_2)...A^{a_{N-1}}(\alpha;\sigma_{N-1}) \cr
& \ \times (p_1 j_z^c(\sigma_1)+p_1 j_z^c(\sigma_N)+\bar p_1 j_{\bar z}^c(\sigma_1)+\bar p_1 j_{\bar z}^c(\sigma_N))
P.V. \frac{1}{\sigma_1-\sigma_N-2\pi}{f^{a_1 a_N}}_c
t_{a_1}...t_{a_N}
\end{align}
We can perform the integral over the free coordinate $\sigma_1$ or $\sigma_N$. We do not find any divergence.

\paragraph{Second-order pole.}
Now we consider the contribution of the second-order pole in the OPE between $A^{a_1}(\alpha;\sigma_1)$ and $A^{a_N}(\alpha;\sigma_N)$ (see Figure \ref{div2M}(b)). We obtain:
\begin{align}
& \int_{2\pi>\sigma_1>...>\sigma_N>0}d\sigma_1...d\sigma_N A^{a_2}(\alpha;\sigma_2)...A^{a_{N-1}}(\alpha;\sigma_{N-1}) 
p_2 P.V. \frac{1}{(\sigma_1-\sigma_N-2\pi)^2}\kappa^{a_1 a_N}
t_{a_1}...t_{a_N}
\end{align}
The integral over the coordinates $\sigma_1$ and $\sigma_N$ leads to:
\be \int_{\sigma_2}^{2\pi} d\sigma_1 \int_0^{\sigma_{N-1}}d\sigma_N P.V. \frac{1}{(\sigma_1-\sigma_N-2\pi)^2}
= - \log \epsilon + \mathrm{finite} \ee
This produces a new logarithmic divergence in the monodromy matrix, that we have to add to the one computed in \eqref{logDivO1}.
So at first order in the perturbative expansion the divergences appearing in the monodromy matrix add up to:
\be\label{MdivO1} p_2 \log \epsilon (C^{(2)} \Omega(\alpha) - \kappa^{ab} t_a \Omega(\alpha) t_b) \ee
Next we consider the new divergences in the monodromy matrix arising at second-order.

\paragraph{OPEs between distinct pairs of connections.}
First we consider the case where two OPEs are taken between two distinct pairs of connections.
From the previous analysis we deduce that these terms lead to the divergences:
\be\label{MdivO2} \frac{1}{2}(p_2 \log \epsilon)^2 ((C^{(2)})^2 \Omega(\alpha) - 2 C^{(2)} \kappa^{ab} t_a \Omega(\alpha) t_b + \kappa^{ab} \kappa^{cd} t_a t_c \Omega(\alpha) t_d t_b) \ee

\paragraph{Triple OPEs.}
The new triple OPE that lead to divergences are the following: we take the first OPE between one connection $A^{a_i}(\alpha;\sigma_i)$ and one of the endpoint connections $A^{a_1}(\alpha;\sigma_1)$ or $A^{a_N}(\alpha;\sigma_N)$. Then we take the OPE between the currents evaluated near the endpoint with the remaining endpoint connection and isolate the contribution of the second-order pole (see Figure \ref{div2M}(c) and \ref{div2M}(d)). The sum of these terms give:
\begin{align}
& \int_{2\pi>\sigma_1>...>\sigma_N>0}d\sigma_1...d\sigma_N A^{a_2}(\alpha;\sigma_2)...A^{a_{i-1}}(\alpha;\sigma_{i-1})A^{a_{i+1}}(\alpha;\sigma_{i+1})...A^{a_{N-1}}(\alpha;\sigma_{N-1}) \cr
& \times (p_1 q_2 + \bar p_1 r_2) 
\left( P.V. \frac{1}{\sigma_1-\sigma_i} P.V. \frac{1}{(\sigma_1-\sigma_N-2\pi)^2} {f^{a_1 a_i}}_c \kappa^{c a_N} \right. \cr
& \left. \qquad + P.V. \frac{1}{\sigma_N-\sigma_i} P.V. \frac{1}{(\sigma_1-\sigma_N-2\pi)^2} {f^{a_N a_i}}_c \kappa^{c a_1}\right)
t_{a_1}...t_{a_N}
\end{align}
We can compute the integrals over $\sigma_1$ and $\sigma_N$:
\begin{align} 
& \int_{\sigma_2}^{2\pi} d\sigma_1 \int_0^{\sigma_{N-1}}d\sigma_N P.V. \frac{1}{\sigma_1-\sigma_i} P.V. \frac{1}{(\sigma_1-\sigma_N-2\pi)^2}
= - \frac{\log \epsilon}{2(2\pi-\sigma_j)} + \mathrm{finite} \cr
& \int_{\sigma_2}^{2\pi} d\sigma_1 \int_0^{\sigma_{N-1}}d\sigma_N P.V. \frac{1}{\sigma_N-\sigma_i} P.V. \frac{1}{(\sigma_1-\sigma_N-2\pi)^2}
= - \frac{\log \epsilon}{2(2\pi-\sigma_j)} + \mathrm{finite}
\end{align}
We deduce that the divergent piece vanishes since:
\be {f^{a_1 a_i}}_c \kappa^{c a_N} + {f^{a_N a_i}}_c \kappa^{c a_1} = 0 \ee

\paragraph{Renormalization of the monodromy matrix.}
Up to second order in perturbation theory the divergences appearing in the monodromy matrix are given by \eqref{MdivO1} and \eqref{MdivO2}.
These divergences can be canceled by a wave-function renormalization of the monodromy matrix:
\begin{align}
\Omega(\alpha) \to \Omega_{ren}(\alpha) & = \Omega(\alpha) -  p_2 \log \epsilon \kappa^{ab} t_a [t_b, \Omega(\alpha)] \cr
& \quad + \frac{1}{2}(p_2 \log \epsilon)^2 \kappa^{ab} \kappa^{cd} t_a t_c [t_b,[t_d, \Omega(\alpha)]] + \mathcal{O}(f^6)
\end{align}
where we used in particular that $\kappa^{ab} t_a t_c t_b = \kappa^{ab} t_a t_b t_c$.
Notice that the operator $\Omega_1(\alpha)$ is not renormalized, since $\kappa^{ab} t_a [t_b, \Omega_1(\alpha)] = 0$. 
This implies in particular that the conserved (local) charge associated with the global symmetry $G_L$ is not renormalized, as expected.

\paragraph{The quantum transfer matrix.}
Taking the trace of the monodromy matrix we obtain the transfer matrix. From the previous discussion it appears that the transfer matrix is completely free of divergences, at least up to second order in perturbation theory. Accordingly there is no need to renormalize the transfer matrix.

This is in sharp contrast with what happens in WZNW models on generic groups. 
In that case the perturbative divergences in the transfer matrix cancel only for special values of the spectral parameter \cite{Bachas:2004sy}.
Nevertheless our result can be anticipated from the analysis of \cite{Bachas:2004sy}, at least at the WZNW points. Indeed it was argued in \cite{Bachas:2004sy}, using a different regularization scheme, that all divergences in the transfer matrix are proportional to the dual Coxeter number of the group.


\section{Fusion of line operators}\label{fusion}

In this section we consider the fusion of two transition matrices. 
The fusion is the process of bringing the integration contours of two line operators on top of each other.
Quantum effects play an important role in this process.

The problem of fusing line operators was previously discussed in the literature in different contexts. 
In \cite{Bachas:2007td} the fusion of conformal interfaces was elucidated in the $c=1$ CFT. 
The fusion of line operators for the pure-spinor string in $AdS_5 \times S^5$
was computed at first-order in perturbation theory in \cite{Mikhailov:2007eg}. 

Let us consider two transition matrices $T^{b+i\epsilon,a+i\epsilon}_R(\alpha)$ and $T^{d,c}_{R'}(\beta)$. 
The contour of the first (resp. second) one lies at constant time $\tau=\epsilon>0$ (resp. $\tau=0$). 
These matrices can be taken in different representations $R$ and $R'$, with different values of the spectral parameter $\alpha$ and $\beta$. 
The fusion of these two transition matrices is defined as:
\be\label{defFusion} \lim_{\epsilon \to 0^+} T^{b+i\epsilon,a+i\epsilon}_R(\alpha)T^{d,c}_{R'}(\beta) \ee
If the intervals $[a,b]$ and $[c,d]$ do not overlap, this process is trivial.
In the following we assume that the overlap of these intervals is non-zero.
For the time being we also assume that the endpoints of the intervals do not coincide; this assumption will be relaxed in section \ref{singularFusion}.
As the distance between the two contours goes to zero, the OPE between two connections $A_R(\alpha;\sigma+i\epsilon)$ and $A_{R'}(\beta;\sigma')$ integrated on the first and on the second contour becomes singular.
For instance let us consider the holomorphic pole in this OPE.
We can rewrite it as:
\begin{align}  \frac{1}{\sigma+i \epsilon - \sigma'} 
& =  \frac{1}{2} \left( \frac{1}{\sigma+i \epsilon - \sigma'} +  \frac{1}{\sigma-i \epsilon - \sigma'} \right)
+ \frac{1}{2} \left( \frac{1}{\sigma+i \epsilon - \sigma'} -  \frac{1}{\sigma-i \epsilon - \sigma'} \right) \cr
& = P.V.\ \frac{1}{\sigma - \sigma'} - i \frac{\epsilon}{(\sigma-\sigma')^2 + \epsilon^2}
\end{align}
The last term is actually a regularization of the delta-function:
\begin{align}
& \delta_\epsilon(\sigma-\sigma') \equiv \frac{1}{\pi}\frac{\epsilon}{(\sigma-\sigma')^2 + \epsilon^2} 
\qquad ; \qquad
 \lim_{\epsilon \to 0^+} \delta_\epsilon(\sigma-\sigma') = \delta(\sigma-\sigma') 
\end{align}
We can perform a similar manipulation for all singularities appearing in the OPE between the two connections $A_R(\alpha;\sigma+i\epsilon)$ and $A_{R'}(\beta;\sigma')$. These singularities are rewritten as%
\footnote{The regularized delta-function in the third line is not exactly the same one as in the first two lines. However to keep the formulas simple we will adopt the same notations for both regularizations of the delta-function.}%
:
\begin{align}
&\frac{1}{(\sigma \pm i \epsilon - \sigma')^2} = P.V. \frac{1}{(\sigma-\sigma')^2} \pm i \pi \delta'_\epsilon(\sigma-\sigma') \cr
&\frac{1}{\sigma \pm i \epsilon - \sigma'} = P.V. \frac{1}{\sigma-\sigma'} \mp i \pi \delta_\epsilon(\sigma-\sigma')  \cr
&\frac{\sigma \mp i \epsilon - \sigma'}{(\sigma \pm i \epsilon - \sigma')^2} = P.V. \frac{1}{\sigma-\sigma'} \mp i \pi \delta_\epsilon(\sigma-\sigma')  
\end{align}
The first term in the previous expressions, the principal value piece, is the regularized singularity that appears in the OPE of two equal-time connections, according to the regularization prescription discussed in section \ref{regularization}. 
So this term is understood as a potential divergence in the double-line operator $T_R^{b,a}(\alpha)  T_{R'}^{d, c}(\beta)$ (with both contours lying at equal time) of the type studied in section \ref{renormalization}.
On the other hand the second term, the regularized delta-function, is a contribution that is specific to the process of fusion. 
Once integrated upon, it gives a finite correction to the classical fusion of line operators. 
These are the corrections we will compute in this section.
 
As before a computation at order $f^{2p}$ involve $p$ OPEs. 
Notice that the regularized delta-function $\delta_\epsilon$ changes sign if we flip the sign of $\epsilon$.
This implies that the quantum corrections associated with fusion that involve an odd number of OPEs contribute to the commutator of the transitions matrices. On the other hand computations involving an even numbers of OPEs contribute to the symmetric product of the transition matrices.


\subsection{First-order corrections}\label{fusionOrder1}

In this section we compute the first-order quantum effects in the fusion of transition matrices.

\paragraph{The relevant OPEs.}
Since we are interested in the quantum corrections associated to the fusion of two line operators, we isolate the delta-function terms in the connection-connection OPE:
\begin{align}\label{OPEAA'delta}
(1-P.V.)& A_{R}(\alpha;\sigma + i \epsilon) A_{R'}(\beta;\sigma') 
=  s \delta'_\epsilon(\sigma - \sigma') \cr
& +  \left[ A_{R}(\alpha;\sigma ),\frac{r+s}{2} \right] \delta_\epsilon(\sigma - \sigma')
+  \left[A_{R'}(\beta;\sigma ),\frac{r-s}{2}\right] \delta_\epsilon(\sigma - \sigma')
\end{align}
The matrices $r$ and $s$ are defined in equation \eqref{rsMat}.
The sub-leading singularities in the current-current OPEs \eqref{jjOPEs} give a vanishing contribution in the limit where $\epsilon$ goes to zero.
Indeed the previous OPE is essentially equal to the commutator \eqref{[A,A']v1}.
All coefficients in the OPE \eqref{OPEAA'delta} follow from the currents two- and three-points functions that were computed from first principles to all orders in $f^2$ in \cite{Ashok:2009xx}.
So the OPE \eqref{OPEAA'delta} is exact to all orders in $f^2$.
As long as we are computing at first order in $f^2$, the precise points at which the connections on the right-hand side are evaluated are not important.

\begin{figure}
\centering
\includegraphics[scale=0.55]{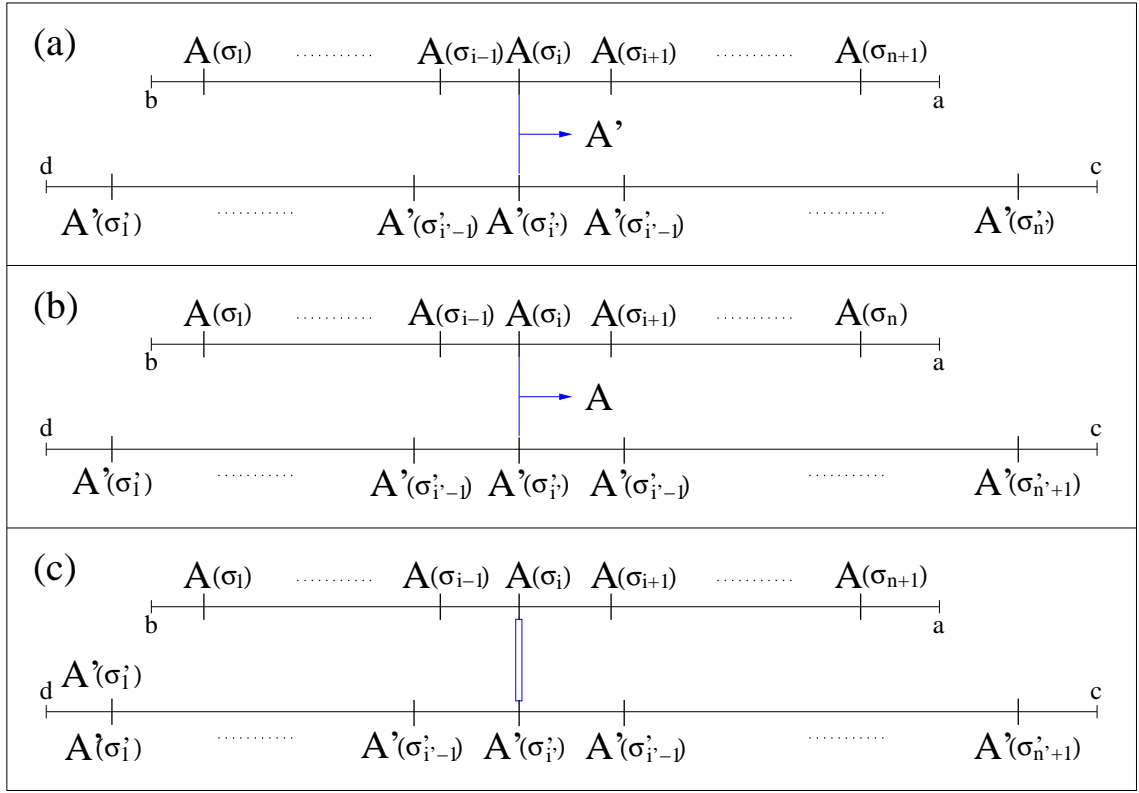}
\caption{Fusion at first-order: three different contributions to the same term in \eqref{expansionInts}. \label{fus1}}
\end{figure}

\paragraph{Computation of the first-order corrections.}
We consider the fusion of two transition matrices:
\be\label{limTT} \lim_{\epsilon \to 0^+} P \exp \left( -\int_{a+i \epsilon}^{b+i \epsilon} d\sigma  A_{R}(\alpha;\sigma) \right) 
P \exp \left( -\int_{c}^{d} d\sigma'  A_{R'}(\beta;\sigma') \right)\ee
To perform the computation we expand the exponential as in equation \eqref{expansionT}. Since the right-hand side of the OPE \eqref{OPEAA'delta} is written in terms of the connections and the constant matrices $r$ and $s$ only, we expect that the result of the computation can be written in the schematic form:
\be\label{expansionInts} \sum_{n,n'=0}^\infty \# \left(\int A_{R}(\alpha)\right)^n \left(\int A_{R'}(\beta)\right)^{n'} \ee
Let us compute the coefficient of a generic term in \eqref{expansionInts}, with $n$ copies of the connection $A_R$ and $n'$ copies of the connection $A_{R'}$.
At order $f^2$, we identify three different contributions to this term (see Figure \ref{fus1}):
\begin{itemize}
	\item From $T^{b+i\epsilon,a+i\epsilon}_{R,n+1} T^{d,c}_{R',n'}$, isolating the contribution of the first-order singularity multiplying $A_{R'}$ in the OPE \eqref{OPEAA'delta}.
	\item From $T^{b+i\epsilon,a+i\epsilon}_{R,n} T^{d,c}_{R',n'+1}$, isolating the contribution of the first-order singularity multiplying $A_{R}$ in the OPE \eqref{OPEAA'delta}.
	\item From $T^{b+i\epsilon,a+i\epsilon}_{R,n+1} T^{d,c}_{R',n'+1}$, isolating the contribution of the second-order singularity in the OPE \eqref{OPEAA'delta}.
\end{itemize}
To simplify the expressions in the following computations, we introduce the shortened notations:
\begin{align} & \left\lfloor \int_x^y A \right\rceil^n \equiv \int_{y>\sigma_1>...>\sigma_n>x} A_R(\alpha;\sigma_1)... A_R(\alpha;\sigma_n) \cr
& \left\lfloor \int_x^y A' \right\rceil^{n'} \equiv \int_{y>\sigma_1>...>\sigma_{n'}>x} A_{R'}(\beta;\sigma_1)... A_{R'}(\beta;\sigma_{n'})
\end{align}
The first contribution coming from $T^{b+i\epsilon,a+i\epsilon}_{R,n+1} T^{d,c}_{R',n'}$ (see Figure \ref{fus1}(a)) reads:
\begin{align}\label{fusionTTO1Term1}
(-)^{n+n'+1} \sum_{i=0}^{n} \sum_{i'=1}^{n'}  \int_{[a,b]\cap[c,d]} d\sigma 
 \left\lfloor \int_\sigma^b A \right\rceil^i   \left\lfloor \int_\sigma^d A' \right\rceil^{i'-1}
\left[A_{R'}(\beta;\sigma),\frac{r-s}{2}\right]
  \left\lfloor \int_a^\sigma A \right\rceil^{n-i}   \left\lfloor \int_c^\sigma A' \right\rceil^{n'-i'}
\end{align}
The second contribution coming from $T^{b+i\epsilon,a+i\epsilon}_{R,n} T^{d,c}_{R',n'+1}$ (see Figure \ref{fus1}(b)) reads:
\begin{align}\label{fusionTTO1Term2}
(-)^{n+n'+1} \sum_{i=1}^{n} \sum_{i'=0}^{n'}  \int_{[a,b]\cap[c,d]} d\sigma 
 \left\lfloor \int_\sigma^b A \right\rceil^{i-1}   \left\lfloor \int_\sigma^d A' \right\rceil^{i'}
\left[A_{R}(\alpha;\sigma),\frac{r+s}{2}\right]
  \left\lfloor \int_a^\sigma A \right\rceil^{n-i}   \left\lfloor \int_c^\sigma A' \right\rceil^{n'-i'}
\end{align}
Eventually the third contribution coming from $T^{b+i\epsilon,a+i\epsilon}_{R,n+1} T^{d,c}_{R',n'+1}$ (see Figure \ref{fus1}(c)) reads:
\begin{align}\label{fusionTTO1Term3}
(-)^{n+n'+2} & \sum_{i=1}^{n+1}\sum_{i'=1}^{n'+1}
\int_{\genfrac{}{}{0pt}{}{b>\sigma_1>...>\sigma_{n+1}>a}{d>\sigma'_1>...>\sigma'_{n'+1}>c}}
A_{R}(\alpha;\sigma_1)...A_{R}(\alpha;\sigma_{i-1})A_{R'}(\beta;\sigma'_1)...A_{R'}(\beta;\sigma'_{i'-1}) \cr
& \times s \delta'_\epsilon(\sigma_i-\sigma'_{i'})
A_{R}(\alpha;\sigma_{i+1})...A_{R}(\alpha;\sigma_{n+1})A_{R'}(\beta;\sigma'_{i'+1})...A_{R'}(\beta;\sigma'_{n'+1})
\end{align}
When we add these three contributions non-trivial simplifications occur.
Details about this computation are given in appendix \ref{CompFusionO1}.
Once the dust has settled we get:
\begin{align}\label{fusionf2nn'}
(-)^{n+n'}& \left( \chi(b;c,d) \sum_{i'=0}^{n'} \left\lfloor \int_{b}^d A' \right\rceil^{i'} \frac{r+s}{2} \left\lfloor \int_{a}^b A \right\rceil^{n} \left\lfloor \int_{c}^b A' \right\rceil^{n'-i'}\right. \cr
&-\chi(a;c,d) \sum_{i'=0}^{n'}  \left\lfloor \int_{a}^b A \right\rceil^{n} \left\lfloor \int_{a}^d A' \right\rceil^{i'}\frac{r+s}{2}  \left\lfloor \int_{c}^a A' \right\rceil^{n'-i'} \cr
&+\chi(d;a,b) \sum_{i=0}^{n} \left\lfloor \int_{d}^b A \right\rceil^{i} \frac{r-s}{2} \left\lfloor \int_{a}^d A \right\rceil^{n-i} \left\lfloor \int_{c}^d A' \right\rceil^{n'} \cr
&\left. -\chi(c;a,b) \sum_{i=0}^{n'}\left\lfloor \int_{c}^b A \right\rceil^{i} \left\lfloor \int_{c}^d A' \right\rceil^{n'} \frac{r-s}{2} \left\lfloor \int_{a}^c A \right\rceil^{n-i}  \right)
\end{align}
\paragraph{Fusion of transition matrices at first-order.}
Using the previous result \eqref{fusionf2nn'} we can perform the sum over the integers $n$ and $n'$ in equation \eqref{expansionInts} to get a compact expression for the fusion of the transition matrices:
\begin{align}\label{fusionf2}
\lim_{\epsilon \to 0^+}&  T^{b+i\epsilon,a+i\epsilon}_R(\alpha)T^{d,c}_{R'}(\beta)
 =  T^{b,a}_R(\alpha)T^{d,c}_{R'}(\beta) \cr
&+ \chi(b;c,d) T^{d,b}_{R'}(\beta) \frac{r+s}{2} T^{b,a}_R(\alpha)T^{b,c}_{R'}(\beta) - \chi(a;c,d) T^{b,a}_R(\alpha) T^{d,a}_{R'}(\beta) \frac{r+s}{2} T^{b,c}_{R'}(\beta) \cr
& + \chi(d;a,b) T^{b,d}_R(\alpha) \frac{r-s}{2} T^{d,a}_R(\alpha) T^{d,c}_{R'}(\beta) - \chi(c;a,b) T^{b,c}_R(\alpha) T^{d,c}_{R'}(\beta) \frac{r-s}{2} T^{c,a}_R(\alpha) \cr
& + \mathcal{O}(f^4)
\end{align}
We recover the formula \eqref{poissonTT} previously obtained in \cite{Maillet:1985ek} using Hamiltonian methods (see appendix \ref{Maillet}).


\subsection{Second-order corrections}\label{fusionOrder2}

Now we will compute order-$f^4$ corrections to the fusion of two transition matrices. 
This implies that we have to take two OPEs. 
These OPEs can be taken between two distinct pairs of connections. 
But we can also take the OPE of two connections, and then take the OPE of the resulting currents with a third connection; in this section we refer to this process as a triple collision.

In order to compute the second-order corrections we can start from the first-order result \eqref{fusionf2} we get previously and perform a second OPE between the remaining connections.
This is the first computation we will perform in this paragraph. 
This method is legitimate when the two OPEs are taken between two distinct pairs of connexions.
However it is not accurate when considering triple collisions.
In a second computation we will evaluate the additional corrections resulting from the triple collisions.

\paragraph{Two successive OPEs.}
In the previous section we considered transition matrices on generic overlapping intervals $[a,b]$ and $[c,d]$.
At second order the equations become heavy if one chooses to work with generic intervals.
So we will work in a particular configuration: we assume that $c<a<b<d$, and describe the computations in this case.
At the end we will also give the results corresponding to the other possible configurations of overlapping intervals.

We consider the order-$f^2$ terms computed in \eqref{fusionf2}. We lift back the contour integration for the connections $A_R(\alpha)$ up to an infinitesimal time $\tau=\epsilon>0$ and we perform a new OPE between two connections $A_R(\alpha)$ and $A_{R'}(\beta)$.
Some OPEs taken between connections separated by an additional constant matrix $r\pm s$ do contribute.
In this case the OPE \eqref{OPEAA'delta} becomes:
\begin{align}\label{OPEAQA'delta}
&(1-P.V.) A_{R}(\alpha;\sigma + i \epsilon) (r\pm s) A_{R'}(\beta;\sigma') 
=  s (r \pm s) \delta'_\epsilon(\sigma - \sigma') +... \cr
&(1-P.V.) A_{R'}(\beta;\sigma') (r\pm s) A_{R}(\alpha;\sigma + i \epsilon) 
=  s (r \pm s) \delta'_\epsilon(\sigma - \sigma') +...
\end{align}
where we used:
\be\label{TQT=Q2} \kappa^{ab} t_a^R (\kappa^{cd} t_c^R t_d^{R'}) t_b^{R'} = \kappa^{ab} t_a^R t_b^{R'} \kappa^{cd} t_c^R t_d^{R'}  \ee
The ellipses in \eqref{OPEAQA'delta} contain only terms proportional to the delta function that will drop out of the following computations.

From the OPEs \eqref{OPEAA'delta} and \eqref{OPEAQA'delta}, we expect that we can once again write the result of the computation in the form \eqref{expansionInts}.
We compute the term in \eqref{expansionInts} involving $n$ connections $A_R(\alpha)$ and $n'$ connections $A_{R'}(\beta)$.
We identify three contributions to this term, that we evaluate next.

The first contribution comes from the order-$f^2$ term in \eqref{fusionf2} with $n+1$ connections $A_R(\alpha)$ and $n'$ connections $A_{R'}(\beta)$.
So the relevant starting point is (see \eqref{fusionf2nn'}):
\begin{align}
(-)^{n+n'+1}\sum_{i'=0}^{n'}  & \left(  \left\lfloor \int_{b}^d A' \right\rceil^{i'} \frac{r+s}{2} \left\lfloor \int_{a}^b A \right\rceil^{n+1} \left\lfloor \int_{c}^b A' \right\rceil^{n'-i'}
\right. \cr
&  \left. 
- \left\lfloor \int_{a}^b A \right\rceil^{n+1} \left\lfloor \int_{a}^d A' \right\rceil^{i'}\frac{r+s}{2}  \left\lfloor \int_{c}^a A' \right\rceil^{n'-i'}  \right)
\end{align}
We take one further OPE between two connections $A_R(\alpha)$ and $A_{R'}(\beta)$, and isolate the term multiplying $A_{R'}(\beta)$ in this OPE.
We obtain:
\begin{align}
(-)^{n+n'+1}\sum_{i'=0}^{n'}  & \sum_{j=0}^{n}  \int_{[a,b]\cap[c,d]} d\sigma  \left(  
 \left\lfloor \int_{b}^d A' \right\rceil^{i'} \frac{r+s}{2}  \sum_{j'=1}^{n'-i'}
\left\lfloor \int_\sigma^b A \right\rceil^j   \left\lfloor \int_\sigma^b A' \right\rceil^{j'-1} \right. \cr
& \qquad\qquad \times
\left[A_{R'}(\beta;\sigma),\frac{r-s}{2}\right]
\left\lfloor \int_a^\sigma A \right\rceil^{n-j}   \left\lfloor \int_c^\sigma A' \right\rceil^{n'-i'-j'} \cr
& -  \sum_{j'=1}^{i'} 
\left\lfloor \int_\sigma^b A \right\rceil^j   \left\lfloor \int_\sigma^d A' \right\rceil^{j'-1} 
\left[A_{R'}(\beta;\sigma),\frac{r-s}{2}\right] \cr
& \qquad\qquad \times \left.
\left\lfloor \int_a^\sigma A \right\rceil^{n-j}   \left\lfloor \int_a^\sigma A' \right\rceil^{i'-j'} 
\frac{r+s}{2}  \left\lfloor \int_{c}^a A' \right\rceil^{n'-i'}  \right)
\end{align}
The second contribution comes from the order-$f^2$ term in \eqref{fusionf2} with $n$ connections $A_R(\alpha)$ and $n'+1$ connections $A_{R'}(\beta)$.
So we start from (see \eqref{fusionf2nn'}):
\begin{align}
(-)^{n+n'+1}\sum_{i'=0}^{n'+1}  & \left(  \left\lfloor \int_{b}^d A' \right\rceil^{i'} \frac{r+s}{2} \left\lfloor \int_{a}^b A \right\rceil^{n} \left\lfloor \int_{c}^b A' \right\rceil^{n'+1-i'}
\right. \cr
&  \left. 
- \left\lfloor \int_{a}^b A \right\rceil^{n} \left\lfloor \int_{a}^d A' \right\rceil^{i'}\frac{r+s}{2}  \left\lfloor \int_{c}^a A' \right\rceil^{n'+1-i'}  \right)
\end{align}
We take one further OPE between two connections $A_R(\alpha)$ and $A_{R'}(\beta)$, and isolate the term multiplying $A_{R}(\alpha)$ in this OPE.
We obtain:
\begin{align}
(-)^{n+n'+1}\sum_{i'=0}^{n'}  & \sum_{j=1}^{n}  \int_{[a,b]\cap[c,d]} d\sigma  \left(  
 \left\lfloor \int_{b}^d A' \right\rceil^{i'} \frac{r+s}{2}  \sum_{j'=1}^{n'-i'}
\left\lfloor \int_\sigma^b A \right\rceil^{j-1}   \left\lfloor \int_\sigma^b A' \right\rceil^{j'} \right. \cr
& \qquad\qquad \times
\left[A_{R}(\alpha;\sigma),\frac{r+s}{2}\right]
\left\lfloor \int_a^\sigma A \right\rceil^{n-j}   \left\lfloor \int_c^\sigma A' \right\rceil^{n'-i'-j'} \cr
& -  \sum_{j'=1}^{i'} 
\left\lfloor \int_\sigma^b A \right\rceil^{j-1}   \left\lfloor \int_\sigma^d A' \right\rceil^{j'} 
\left[A_{R}(\alpha;\sigma),\frac{r+s}{2}\right] \cr
& \qquad\qquad \times \left.
\left\lfloor \int_a^\sigma A \right\rceil^{n-j}   \left\lfloor \int_a^\sigma A' \right\rceil^{i'-j'} 
\frac{r+s}{2}  \left\lfloor \int_{c}^a A' \right\rceil^{n'-i'}  \right)
\end{align}
The third contribution comes from the order-$f^2$ term in \eqref{fusionf2} with $n+1$ connections $A_R(\alpha)$ and $n'+1$ connections $A_{R'}(\beta)$.
So we start from (see \eqref{fusionf2nn'}):
\begin{align}
(-)^{n+n'+2}\sum_{i'=0}^{n'+1}  & \left(  \left\lfloor \int_{b}^d A' \right\rceil^{i'} \frac{r+s}{2} \left\lfloor \int_{a}^b A \right\rceil^{n+1} \left\lfloor \int_{c}^b A' \right\rceil^{n'+1-i'}
\right. \cr
&  \left. 
- \left\lfloor \int_{a}^b A \right\rceil^{n+1} \left\lfloor \int_{a}^d A' \right\rceil^{i'}\frac{r+s}{2}  \left\lfloor \int_{c}^a A' \right\rceil^{n'+1-i'}  \right)
\end{align}
We take one further OPE between two connections $A_R(\alpha)$ and $A_{R'}(\beta)$, and isolate the term multiplying the identity in this OPE.
We obtain:
\begin{align}\label{fusionTTO2Term3}
 (-)^{n+n'+2} & \left(
 \sum_{i'=1}^{n'+1} \int_a^b d\sigma \int_b^d d\sigma'  \left\lfloor \int_{\sigma'}^d A' \right\rceil^{i'-1} \frac{r+s}{2} s \delta'_\epsilon(\sigma-\sigma')
\left\lfloor \int_{a}^\sigma A \right\rceil^{n} \left\lfloor \int_{c}^b A' \right\rceil^{n'+1-i'} \right. \cr
& +  \sum_{i'=0}^{n'} \left\lfloor \int_{b}^d A' \right\rceil^{i'} \frac{r+s}{2} 
\sum_{j=1}^{n+1} \sum_{j'=1}^{n'+1-i'}
\int_{\genfrac{}{}{0pt}{}{b>\sigma_1>...>\sigma_{n+1}>a}{b>\sigma'_1>...>\sigma'_{n'+1-i'}>c}} 
A_{R}(\alpha;\sigma_1)...A_{R}(\alpha;\sigma_{j-1}) \cr
& \quad \times 
 A_{R'}(\beta;\sigma'_1)...A_{R'}(\beta;\sigma'_{j'-1})
s \delta'_\epsilon(\sigma_j-\sigma'_{j'}) A_{R}(\alpha;\sigma_{j+1})...A_{R}(\alpha;\sigma_{n+1}) \cr
& \quad \times 
A_{R'}(\beta;\sigma'_{j'+1})...A_{R'}(\beta;\sigma'_{n'+1-i'}) \cr
& -  \sum_{i'=1}^{n'+1} \int_a^b d\sigma \int_c^a d\sigma' \left\lfloor \int_{a}^d A' \right\rceil^{i'} \left\lfloor \int_\sigma^b A \right\rceil^{n} 
\frac{r+s}{2} s \delta'_\epsilon(\sigma-\sigma')\left\lfloor \int_{c}^a A' \right\rceil^{n'+1-i'} \cr
& -  \sum_{i'=0}^{n'} \sum_{j=1}^{n+1} \sum_{j'=1}^{i'}
\int_{\genfrac{}{}{0pt}{}{b>\sigma_1>...>\sigma_{n+1}>a}{d>\sigma'_1>...>\sigma'_{i'}>a}}
A_{R}(\alpha;\sigma_1)...A_{R}(\alpha;\sigma_{j-1})  \cr
& \quad \times A_{R'}(\beta;\sigma'_1)...A_{R'}(\beta;\sigma'_{j'-1}) s \delta'_\epsilon(\sigma_j-\sigma'_{j'})
A_{R}(\alpha;\sigma_{j+1})...A_{R}(\alpha;\sigma_{n+1}) \cr
& \quad \times \left. A_{R'}(\beta;\sigma'_{j'+1})...A_{R'}(\beta;\sigma'_{i'})
\frac{r+s}{2} \left\lfloor \int_{c}^a A' \right\rceil^{n'+1-i'} \right)
\end{align}
In the previous expression the first term comes from the OPE between the first connection $A_{R}(\alpha;\sigma_{1})$ integrated over the integral $[a,b]$ and the last connection $A_{R'}(\beta;\sigma'_{i'})$ integrated over the interval $[b,d]$. The third term has a similar origin.

Now we add up these three contributions, with an additional factor of $\frac{1}{2}$ since the order in which the two OPEs are performed is irrelevant. 
Simplifications occur following the same pattern as in the first-order computation.
Additional subtleties due to the first and third terms in \eqref{fusionTTO2Term3} are discussed in appendix \ref{CompFusionO2}.
The result takes a rather simple form:
\begin{align}\label{fusionf4nn'}
\frac{1}{2}(-)^{n+n'} & \left(
\sum_{i'=0}^{n'} \left\lfloor \int_{b}^d A' \right\rceil^{i'} \left(\frac{r+s}{2}\right)^2 
\left\lfloor \int_a^b A \right\rceil^{n} \left\lfloor \int_{c}^b A' \right\rceil^{n'-i'} \right.\cr
&-  2\sum_{i'=0}^{n'}\sum_{j'=0}^{n'-i'}\left\lfloor \int_{b}^d A' \right\rceil^{i'}  \frac{r+s}{2} 
\left\lfloor \int_a^b A \right\rceil^{n} \left\lfloor \int_{a}^b A' \right\rceil^{j'} 
\frac{r+s}{2} \left\lfloor \int_{c}^a A' \right\rceil^{n'-i'-j'}  \cr
&\left. + \sum_{i'=0}^{n'} \left\lfloor \int_a^b A \right\rceil^{n} \left\lfloor \int_{a}^d A' \right\rceil^{i'} 
 \left(\frac{r+s}{2}\right)^2  \left\lfloor \int_{c}^a A' \right\rceil^{n'-i'} \right)\end{align}
Adding up the zeroth- and first-order contributions computed previously, we get a compact expression:
\begin{align}\label{fusion2step} T^{d,b}_{R'}(\beta) e^{\frac{r+s}{2}} T^{b,a}_R(\alpha)T^{b,a}_{R'}(\beta) e^{-\frac{r+s}{2}} T^{a,c}_{R'}(\beta)  \end{align}
This result is valid up to terms of order $f^6$. Also it has to be corrected by order-$f^4$ terms resulting from triple collisions.
For different configurations of overlapping intervals, equation \eqref{fusion2step} is modified.
In the generic case it reads:
\begin{align} 
& \chi(b;c,d) \chi(a;c,d) T^{d,b}_{R'}(\beta) e^{\frac{r+s}{2}} T^{b,a}_R(\alpha)T^{b,a}_{R'}(\beta) e^{-\frac{r+s}{2}} T^{a,c}_{R'}(\beta) \cr
& + \chi(d;a,d) \chi(c;a,b) T^{b,d}_R(\alpha) e^{\frac{r-s}{2}} T^{d,c}_R(\alpha)T^{d,c}_{R'}(\beta) e^{-\frac{r-s}{2}} T^{c,a}_R(\alpha)  \cr
& + \chi(d;a,b) \chi(a;c,d) T^{b,d}_R(\alpha) e^{\frac{r-s}{2}} T^{d,a}_R(\alpha)T^{d,a}_{R'}(\beta) e^{-\frac{r+s}{2}} T^{a,c}_{R'}(\beta) \cr
& + \chi(b;c,d) \chi(c;a,b) T^{d,b}_{R'}(\beta) e^{\frac{r+s}{2}} T^{b,c}_R(\alpha)T^{b,c}_{R'}(\beta) e^{-\frac{r-s}{2}} T^{c,a}_R(\alpha)
\end{align}
As mentioned previously this result does not properly take into account the corrections produced by triple collisions.
Next we will evaluate these additional contributions.


\paragraph{Additional corrections from triple collisions.}

To compute the quantum corrections that appear when three connections collide, we first perform the OPE between two connections (first step), and then perform the OPE of the resulting currents with the third connection (second step). 
The point at which the currents appearing in the first OPE are evaluated is important.
Indeed regularized delta-functions only appear in the OPEs between operators evaluated at different time, and only these terms contribute to the quantum corrections associated with fusion.
If we use the OPE \eqref{OPEAA'delta} for the first step of the computation, we implicitly distribute the resulting currents arbitrarily on one or the other contour so that they combine into connections evaluated on these contours.
This is the reason why the previous computation does not take into account properly the triple collisions of OPEs.
Instead of \eqref{OPEAA'delta}, we have to use the following OPE in the first step of the computation:
\begin{align}\label{OPEAA'deltaEpsilon}
(1-P.V.)& A^a(\alpha;\sigma+i \epsilon) A^b(\beta,\sigma') = f^2 \kappa^{ab} a \delta'_\epsilon(\sigma-\sigma')\cr
&+ f^2 i{f^{ab}}_c \left( b_1 j^c_z(\sigma+i \epsilon)+ b_2 j^c_z(\sigma') + \bar b_1 j^c_{\bar z}(\sigma+i \epsilon)+ \bar b_2 j^c_{\bar z}(\sigma') \right) \delta_\epsilon(\sigma-\sigma')
\end{align}
The coefficients can be deduced from equation \eqref{OPEAA'}. It is convenient to write them as:
\begin{align}
&\frac{a}{i\pi} = N_R - N_{R'} \cr
&\frac{b_1}{i\pi} = \frac{1}{2}\left( - \frac{2}{1+\alpha}N_R + \frac{2}{1+\beta}N_{R'}\right) + (1-\eta)^2 \Delta \cr
&\frac{b_2}{i\pi} = \frac{1}{2}\left( - \frac{2}{1+\alpha}N_R + \frac{2}{1+\beta}N_{R'}\right) - (1-\eta)^2 \Delta  \cr
&\frac{\bar b_1}{i\pi} = \frac{1}{2}\left( - \frac{2}{1-\alpha}N_R + \frac{2}{1-\beta}N_{R'}\right) + \eta^2 \Delta  \cr
&\frac{\bar b_2}{i\pi} = \frac{1}{2}\left( - \frac{2}{1-\alpha}N_R + \frac{2}{1-\beta}N_{R'}\right) - \eta^2 \Delta  
\end{align}
where $N_R$ and $N_{R'}$ are given in \eqref{NN'} and $\Delta$ is given by:
\be \Delta = 4 \frac{\beta-\alpha}{(1-\alpha^2)(1-\beta^2)} \ee
%
Of course in the limit $\epsilon \to 0^+$ the OPEs \eqref{OPEAA'delta} and \eqref{OPEAA'deltaEpsilon} are identical. But this limit has to be taken after we perform the second OPE.
In this paragraph we compute the corrections to the previous computation where we started from the first-order result.
So we have to subtract a piece in the corrections associated to the triple collisions that was already accounted for in the previous computation.
This amounts to perform the replacement:
\begin{align}
&b_1 \to \tilde b_1 = b_1 -  \left( -i\pi \frac{2}{1+\alpha} N_R \right)
\qquad ; \qquad
b_2 \to \tilde b_2 = b_2 -  \left( i\pi \frac{2}{1+\beta} N_{R'} \right)\cr
&\bar b_1 \to \tilde{\bar b}_1 = \bar b_1 - \left( -i\pi \frac{2}{1-\alpha} N_R \right)
\qquad ; \qquad
\bar b_2 \to \tilde{\bar b}_2 = \bar b_2 - \left( i\pi \frac{2}{1-\beta} N_{R'} \right)
\end{align}
In the second step of the computation we perform the OPE between the currents in \eqref{OPEAA'deltaEpsilon} and a third connection.
The OPEs between a current evaluated at $\tau=\epsilon$ and a connection $A_{R'}(\beta)$ evaluated at $\tau=0$ are:
\begin{align}
(1-P.V.)& j^a_z(\sigma+i \epsilon) A^b(\beta,\sigma') = f^2 \kappa^{ab} c_\beta \delta'_\epsilon(\sigma-\sigma')
+ f^2 i{f^{ab}}_c \left( d_\beta j^c_z(\sigma) + \bar d_\beta j^c_{\bar z}(\sigma) \right) \delta_\epsilon(\sigma-\sigma')
\cr 
(1-P.V.) & j^a_{\bar z}(\sigma+i \epsilon) A^b(\beta,\sigma') = f^2 \kappa^{ab} e_\beta \delta'_\epsilon(\sigma-\sigma')
+ f^2 i{f^{ab}}_c \left( f_\beta j^c_z(\sigma) + \bar f_\beta j^c_{\bar z}(\sigma) \right) \delta_\epsilon(\sigma-\sigma')
\end{align}
The coefficients can be read from \eqref{jjOPEs}:
\begin{align}
&\frac{c_\beta}{i\pi} = \frac{2}{1+\beta} \eta^2 
\quad ; \quad
\frac{d_\beta}{i\pi} = -\frac{2}{1+\beta}\eta(2-\eta) + \frac{2}{1-\beta}(1-\eta)^2 
\quad ; \quad
\frac{\bar d_\beta}{i\pi} = -\frac{2}{1+\beta}\eta^2 - \frac{2}{1-\beta}\eta^2
\cr
&\frac{e_\beta}{i\pi} = -\frac{2}{1-\beta} (1-\eta)^2 
\quad ; \quad
\frac{f_\beta}{i\pi} = \frac{2}{1+\beta}(1-\eta)^2 + \frac{2}{1-\beta}(1-\eta)^2  
\cr &
\frac{\bar f_\beta}{i\pi} = -\frac{2}{1+\beta}\eta^2 + \frac{2}{1-\beta}(1-\eta)(1+\eta) 
\end{align}
The OPEs between a current evaluated at $\tau=0$ and a connection $A_R(\alpha)$ evaluated at $\tau=\epsilon$ are similar, up to an overall sign and the obvious exchange of $\alpha$ and $\beta$.
The precise point at which the resulting currents are evaluated is not relevant anymore in the last step of the computation.

\begin{figure}
\centering
\includegraphics[scale=0.55]{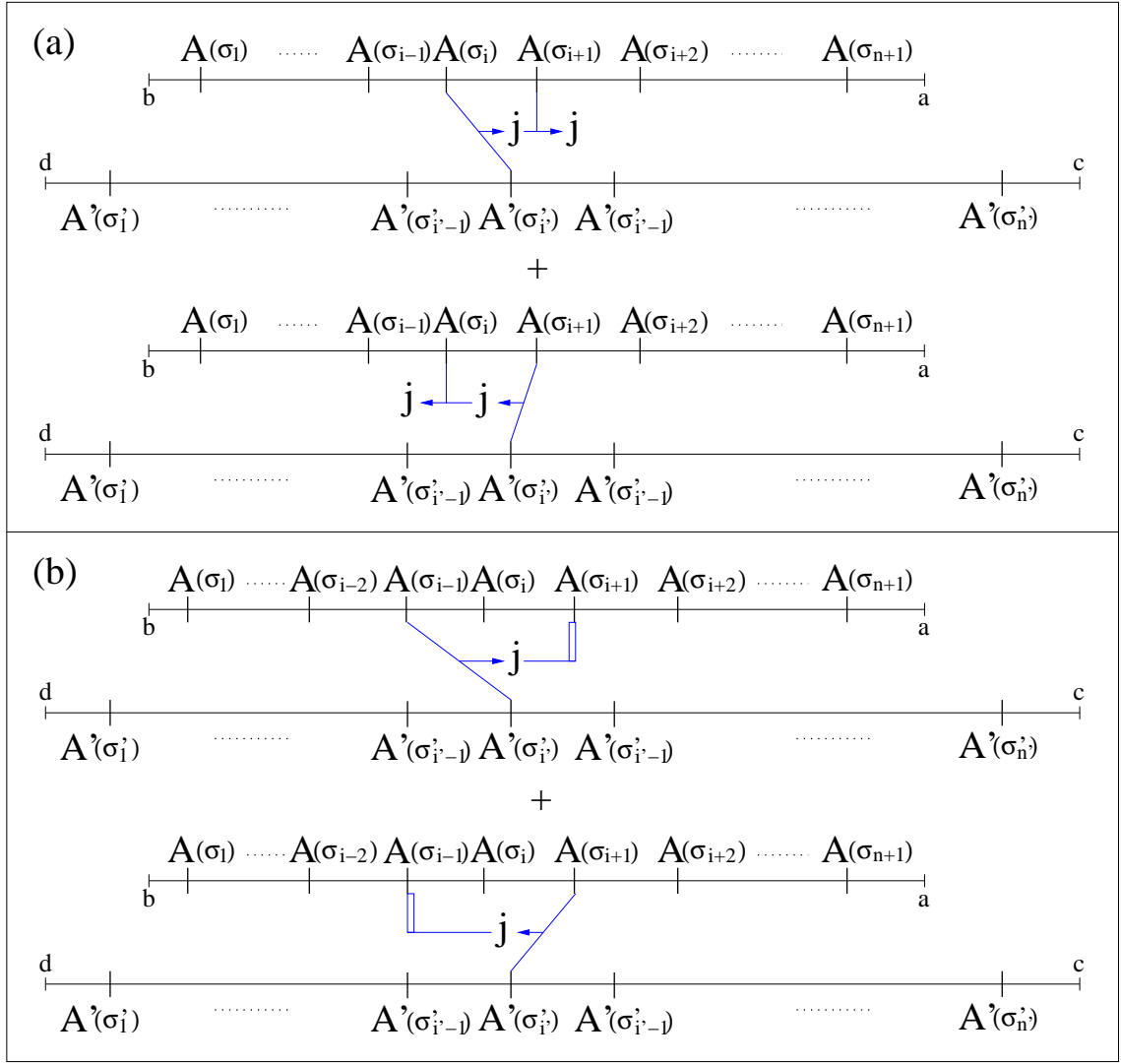}
\caption{Fusion at second-order: examples of triple collisions. \label{fus2}}
\end{figure}

The additional corrections coming from the triple collisions can be written in the form:
\be\label{3colExps} \sum_{n,n'=0}^\infty \# \int j_z \left(\int A_R(\alpha)\right)^n \left(\int A_{R'}(\beta)\right)^{n'} 
+ \sum_{n,n'=0}^\infty \# \int j_{\bar z} \left(\int A_R(\alpha)\right)^n \left(\int A_{R'}(\beta)\right)^{n'} \ee
We will now compute the term in \eqref{3colExps} that contains the $z$-component of the current together with respectively $n$ and $n'$ connections $A_R(\alpha)$ and $A_{R'}(\beta)$.
We identify four different contributions to this term.
The first one (see Figure \ref{fus2}(a)) comes from  $T^{b+i\epsilon,a+i\epsilon}_{R,n+2} T^{d,c}_{R',n'+1}$.
In the second step we take the OPE between a current and a connection $A_R(\alpha)$, and isolate the term multiplying the $z$-component of the current.
We obtain:
\begin{align}
&(-)^{n+n'+3} 2\frac{1}{2}f^4 \sum_{i=0}^{n} \sum_{i'=0}^{n'} \int_{[a,b]\cap[c,d]} d\sigma  
\left\lfloor \int_\sigma^b A \right\rceil^{i} \left\lfloor \int_\sigma^d A' \right\rceil^{i'} \cr
&\qquad \times i{f^{ef}}_g i{f^{gh}}_k j^k_z(\sigma) t_{(h}^R t^R_{e)}t^{R'}_f
\left(-\tilde b_2 d_\alpha - \tilde{\bar b}_2 f_\alpha \right)
\left\lfloor \int_a^\sigma A \right\rceil^{n-i} \left\lfloor \int_c^\sigma A' \right\rceil^{n'-i'} 
\end{align}
The factor of two comes from the two different way of contracting the three connections depicted in Figure \ref{fus2}(a).
The factor of one half comes from the evaluation of the integral over the regularized delta-functions:
\begin{align}
& \int_{\tilde{b}>\sigma_1>\sigma_2>\tilde{a}}d\sigma_1 d\sigma_2 \delta_\epsilon(\sigma_1-\sigma')\delta_\epsilon(\sigma_2-\sigma')
= \frac{1}{2} \chi(\sigma';\tilde{a},\tilde{b})
\end{align}
Notice also that the vanishing of the dual Coxeter number together with the Jacobi identity imply that:
\be {f^{bc}}_d {f^{da}}_e t^R_{a}t^R_{b} = {f^{bc}}_d {f^{da}}_e t^R_{(a}t^R_{b)}\ee
The second contribution comes from $T^{b+i\epsilon,a+i\epsilon}_{R,n+1} T^{d,c}_{R',n'+2}$.
In the second step we take the OPE between a current and a connection $A_{R'}(\beta)$, and isolate the term multiplying the $z$-component of the current.
We obtain:
%
%
\begin{align}
&(-)^{n+n'+3} 2\frac{1}{2}f^4 \sum_{i=0}^{n} \sum_{i'=0}^{n'} \int_{[a,b]\cap[c,d]} d\sigma  
\left\lfloor \int_\sigma^b A \right\rceil^{i} \left\lfloor \int_\sigma^d A' \right\rceil^{i'} \cr
&\qquad \times i{f^{ef}}_g i{f^{gh}}_k j^k_z(\sigma) t^R_{e}t^{R'}_{(h}t^{R'}_{f)}
\left(\tilde b_1 d_\beta + \tilde{\bar b}_1 f_\beta \right)
\left\lfloor \int_a^\sigma A \right\rceil^{n-i} \left\lfloor \int_c^\sigma A' \right\rceil^{n'-i'} 
\end{align}
The third contribution (see Figure \ref{fus2}(b)) comes from $T^{b+i\epsilon,a+i\epsilon}_{R,n+3} T^{d,c}_{R',n'+1}$.
%
In the second step we take the OPE between a current and a connection $A_{R}(\alpha)$, and isolate the term coming with the identity operator in this OPE. The two connections $A_{R}(\alpha)$ involved in this triple OPE have to be separated by exactly one other connection $A_{R}(\alpha)$, else the result vanishes. We extract from this remaining connection the term multiplying the $z$-component of the current.
We obtain:
\begin{align}
&(-)^{n+n'+4} 2\frac{1}{2}f^4 \sum_{i=0}^{n} \sum_{i'=0}^{n'} \int_{[a,b]\cap[c,d]} d\sigma  
\left\lfloor \int_\sigma^b A \right\rceil^{i} \left\lfloor \int_\sigma^d A' \right\rceil^{i'} \cr
&\qquad \times i{f^{ef}}_g i{f^{gh}}_k j^k_z(\sigma) t_{(h}^R t^R_{e)}t^{R'}_f
\left(\tilde b_2 c_\alpha \frac{2}{1+\alpha} + \tilde{\bar b}_2 e_\alpha \frac{2}{1+\alpha}\right)
\left\lfloor \int_a^\sigma A \right\rceil^{n-i} \left\lfloor \int_c^\sigma A' \right\rceil^{n'-i'} 
\end{align}
The factor of $\frac{1}{2}$ comes from the integral over the regularized delta-functions:
\begin{align}
& \int_{\tilde{\sigma}}^{\tilde{b}}d\sigma_1 \int_{\tilde{a}}^{\tilde{\sigma}} d\sigma_2 \int_{\tilde{c}}^{\tilde{d}} d \sigma' \delta_\epsilon(\sigma_1-\sigma')\delta'_\epsilon(\sigma_2-\sigma')
= \frac{1}{2} \chi(\tilde{\sigma};c,d)
\end{align}
The fourth contribution is similar to the third one. It comes from $T^{b+i\epsilon,a+i\epsilon}_{R,n+1} T^{d,c}_{R',n'+3}$.
We obtain:
\begin{align}
&(-)^{n+n'+4} 2\frac{1}{2}f^4 \sum_{i=0}^{n} \sum_{i'=0}^{n'} \int_{[a,b]\cap[c,d]} d\sigma  
\left\lfloor \int_\sigma^b A \right\rceil^{i} \left\lfloor \int_\sigma^d A' \right\rceil^{i'} \cr
&\qquad \times i{f^{ef}}_g i{f^{gh}}_k j^k_z(\sigma) t^R_{e}t^{R'}_{(h}t^{R'}_{f)}
\left(-\tilde b_1 c_\beta\frac{2}{1+\beta} - \tilde{\bar b}_1 e_\beta \frac{2}{1+\beta} \right)
\left\lfloor \int_a^\sigma A \right\rceil^{n-i} \left\lfloor \int_c^\sigma A' \right\rceil^{n'-i'} 
\end{align}
Summing up these four contributions we get:
\begin{align}
&(-)^{n+n'} f^4 \sum_{i=0}^{n} \sum_{i'=0}^{n'} \int_{[a,b]\cap[c,d]} d\sigma  
\left\lfloor \int_\sigma^b A \right\rceil^{i} \left\lfloor \int_\sigma^d A' \right\rceil^{i'} \cr
&\qquad \times i{f^{ef}}_g i{f^{gh}}_k j^k_z(\sigma) 
\left( h_1 t^R_{(e}t^R_{h)}t^{R'}_{f} + h_2 t^R_{e}t^{R'}_{(h}t^{R'}_{f)} \right)
\left\lfloor \int_a^\sigma A \right\rceil^{n-i} \left\lfloor \int_c^\sigma A' \right\rceil^{n'-i'} 
\end{align}
with:
\begin{align}
\frac{h_1}{\pi^2} & = \left( \frac{1}{2} \left(\frac{2}{1+\alpha}N_R + \frac{2}{1+\beta}N_{R'} \right) + (1-\eta)^2 \Delta \right)\frac{2}{1+\alpha}\frac{(1+\alpha-2\eta)^2}{1-\alpha^2} \cr
\frac{h_2}{\pi^2} & = \left( \frac{1}{2} \left(\frac{2}{1+\alpha}N_R + \frac{2}{1+\beta}N_{R'}\right) + (1-\eta)^2 \Delta \right)\frac{2}{1+\beta}\frac{(1+\beta-2\eta)^2}{1-\beta^2} 
\end{align}
Notice that some remarkable simplifications occurred:
\begin{align}
& d_\alpha + \frac{2}{1+\alpha} c_\alpha = \frac{2}{1+\alpha} \frac{(1+\alpha-2\eta)^2}{1-\alpha^2} \cr
& f_\alpha + \frac{2}{1+\alpha} e_\alpha = 0 
\end{align}
Similarly we can compute the term in \eqref{3colExps} that contains the $\bar z$-component of the current together with respectively $n$ and $n'$ connections $A_R(\alpha)$ and $A_{R'}(\beta)$.
We obtain:
\begin{align}
&(-)^{n+n'} f^4 \sum_{i=0}^{n} \sum_{i'=0}^{n'} \int_{[a,b]\cap[c,d]} d\sigma  
\left\lfloor \int_\sigma^b A \right\rceil^{i} \left\lfloor \int_\sigma^d A' \right\rceil^{i'} \cr
&\qquad \times i{f^{ef}}_g i{f^{gh}}_k j^k_{\bar z}(\sigma) 
\left( \bar h_1 t^R_{(e}t^R_{h)}t^{R'}_{f} + \bar h_2 t^R_{e}t^{R'}_{(h}t^{R'}_{f)} \right)
\left\lfloor \int_a^\sigma A \right\rceil^{n-i} \left\lfloor \int_c^\sigma A' \right\rceil^{n'-i'} 
\end{align}
with:
\begin{align}
\frac{\bar h_1}{\pi^2} & = -\left( \frac{1}{2} \left(\frac{2}{1-\alpha}N_R + \frac{2}{1-\beta}N_{R'} \right) + \eta^2 \Delta \right) \frac{2}{1-\alpha}\frac{(1+\alpha-2\eta)^2}{1-\alpha^2} \cr
\frac{\bar h_2}{\pi^2} & = - \left( \frac{1}{2} \left(\frac{2}{1-\alpha}N_R + \frac{2}{1-\beta}N_{R'} \right) + \eta^2 \Delta \right) \frac{2}{1-\beta}\frac{(1+\beta-2\eta)^2}{1-\beta^2} 
\end{align}

\subsection*{Summary: Fusion at second-order}

From the previous computations we deduce the fusion of two transition matrices, up to second order in the $f^2$ expansion:
\begin{align}\label{fusionf4}
\lim_{\epsilon \to 0^+} T^{b+i\epsilon,a+i\epsilon}_R & (\alpha)T^{d,c}_{R'}(\beta) 
=  \chi(b;c,d) \chi(a;c,d) T^{d,b}_{R'}(\beta) e^{\frac{r+s}{2}} T^{b,a}_R(\alpha)T^{b,a}_{R'}(\beta) e^{-\frac{r+s}{2}} T^{a,c}_{R'}(\beta) \cr
&  + \chi(d;a,d) \chi(c;a,b) T^{b,d}_R(\alpha) e^{\frac{r-s}{2}} T^{d,c}_R(\alpha)T^{d,c}_{R'}(\beta) e^{-\frac{r-s}{2}} T^{c,a}_R(\alpha)  \cr
& + \chi(d;a,b) \chi(a;c,d) T^{b,d}_R(\alpha) e^{\frac{r-s}{2}} T^{d,a}_R(\alpha)T^{d,a}_{R'}(\beta) e^{-\frac{r+s}{2}} T^{a,c}_{R'}(\beta) \cr
& + \chi(b;c,d) \chi(c;a,b) T^{d,b}_{R'}(\beta) e^{\frac{r+s}{2}} T^{b,c}_R(\alpha)T^{b,c}_{R'}(\beta) e^{-\frac{r-s}{2}} T^{c,a}_R(\alpha) \cr
&+  f^4  \int_{[a,b]\cap[c,d]} d\sigma  T^{b,\sigma}_R(\alpha)T^{d,\sigma}_{R'}(\beta)
 i{f^{ef}}_g i{f^{gh}}_k 
 \left(
\left( h_1 t^R_{(e}t^R_{h)}t^{R'}_{f} + h_2 t^R_{e}t^{R'}_{(h}t^{R'}_{f)} \right)j^k_z(\sigma) \right.
 \cr
 & \qquad 
+ \left. \left( \bar h_1 t^R_{(e}t^R_{h)}t^{R'}_{f} + \bar h_2 t^R_{e}t^{R'}_{(h}t^{R'}_{f)} \right)j^k_{\bar z}(\sigma) \right) 
T^{\sigma,a}_R(\alpha)T^{\sigma,c}_{R'}(\beta)
\cr &
 + \mathcal{O}(f^6)
\end{align}
The first four terms give a natural generalization of the first-order result \eqref{fusionf2}. The remaining term codes new corrections coming from the simultaneous collision of three connections.

\begin{figure}
\centering
\includegraphics[scale=0.55]{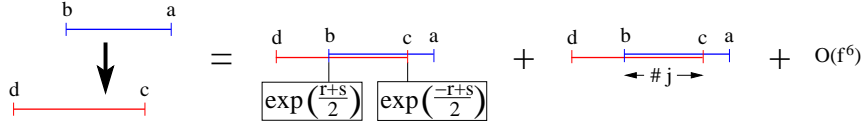}
\caption{Schematic picture of the fusion of two line operators at second order in perturbation theory. \label{summaryFus}}
\end{figure}


\subsection{Fusion of transition matrices with coinciding endpoints}\label{singularFusion}

In the previous analysis we assumed that the integration contours of the line operators we fused had non-coincident endpoints.
Now we consider the case where some endpoints do coincide.
A subtlety arises from the double poles in the current-current OPEs, which gives rise to the derivative of the regularized delta-function in the OPEs \eqref{OPEAA'delta}, \eqref{OPEAA'deltaEpsilon} relevant for fusion.
Technically the problem is that the double integral of the derivative of the delta-function over intervals with coinciding endpoints is not well-defined.
In the Hamiltonian framework fixing this issue requires a somewhat ad hoc regularization procedure \cite{Maillet:1985ek}: one has to perform a symmetric point-splitting to separate the coinciding endpoints
(see appendix \ref{Maillet}).

In our case the integrated delta-functions are naturally regularized with the parameter $\epsilon$ that controls the distance between the two integration contours. 
This regularization leads to an unambiguous result for the fusion of line operators even if the endpoints of the integration contours coincide.
Let us see how this works in a simple example.
We consider the quantity:
\be \lim_{\epsilon \to 0^+} \int_{a+i\epsilon}^{b+i\epsilon}d\sigma A_R(\alpha;\sigma) \int_a^d d\sigma' A_{R'}(\beta;\sigma') \ee
We are interested in the contribution of the double-pole in the OPE between the two connections.
Using \eqref{OPEAA'delta}, we get:
\be \int_{a}^b d\sigma \int_a^d d\sigma' s \delta'_\epsilon(\sigma-\sigma') \ee
We can perform this integral using the explicit form of the regularized delta-function. We obtain:
\begin{align}
\lim_{\epsilon \to 0^+} \frac{1}{\pi} \left( \arctan\left( \frac{b-a}{\epsilon} \right)
-\arctan\left( \frac{d-a}{\epsilon} \right) 
 + \arctan\left( \frac{d-b}{\epsilon} \right) \right)
  = \frac{1}{2} \mathrm{sign}(d-b)
\end{align}
This result can be reproduced from the double integral of the derivative of the (non-regularized) delta-function, given that we use a symmetric point-splitting to separate the coinciding endpoints:
\begin{align}
\frac{1}{2} \left(
 \int_{a+\tilde{\epsilon}}^b d\sigma \int_a^d d\sigma' s \delta'(\sigma-\sigma') 
 +  \int_{a-\tilde{\epsilon}}^b d\sigma \int_a^d d\sigma' s \delta'_\epsilon(\sigma-\sigma') 
 \right) = \frac{1}{2} \mathrm{sign}(d-b)
 \end{align}
where $\tilde{\epsilon}$ is infinitesimal.

The equivalence between the natural regularization we are using and the symmetric point-splitting prescription is generic.
Consequently our method leads to the same results than the Maillet-regularization \cite{Maillet:1985ek}.
Using the symmetric point-splitting prescription it is straightforward to deduce the fusion of line operators with coinciding endpoints from \eqref{fusionf4}.


\subsection{Fusion of monodromy and transfer matrices}

Let us now consider the fusion of two monodromy matrices.
To make use of our previous results we work on the universal cover of the cylinder. 
The contour defining a first monodromy matrix $\Omega_{R'}^{(0)}(\beta)$ is lifted to the interval $[0,2\pi]$.
We consider a second monodromy matrix $\Omega_{R}^{(\epsilon)}(\alpha)$ that we want to fuse with the first one.
We also have to lift the integration contour of the second monodromy matrix to the plane.
There are three different choices with a non-zero overlap with the first contour: $[0,2\pi]$, $[-2\pi,0]$ and $[2\pi,4\pi]$.
To compute the fusion of the monodromy matrices we have to sum the quantum corrections obtained from these three possible lifts \cite{Maillet:1985fn}\cite{Maillet:1985ek}.
Some second-order corrections get contributions from two different lifts at the same time.
The result follows from equation \eqref{fusionf4}:
\begin{align}\label{fusionMM}
\lim_{\epsilon \to 0^+} \Omega_{R}^{(\epsilon)}&(\alpha)\Omega_{R'}^{(0)}(\beta) =
\Omega_{R}(\alpha)\Omega_{R'}(\beta) 
 + \frac{1}{2} \left(r \Omega_{R}(\alpha)\Omega_{R'}(\beta) - \Omega_{R}(\alpha)\Omega_{R'}(\beta)r \right) \cr
& + \frac{1}{2} \kappa^{ab} \tilde s \left( t^{R'}_{b}\Omega_{R}(\alpha) \Omega_{R'}(\beta)t^R_{a} - t^R_{a}\Omega_{R'}(\beta) \Omega_{R}(\alpha)t^{R'}_{b} \right) \cr
& + \frac{1}{8} \left(r^2 \Omega_{R}(\alpha)\Omega_{R'}(\beta) - 2r \Omega_{R}(\alpha)\Omega_{R'}(\beta)r + \Omega_{R}(\alpha)\Omega_{R'}(\beta) r^2 \right) 
\cr 
& - \frac{1}{8} \kappa^{ab} \tilde s \left(- t^{R'}_{b}\Omega_{R}(\alpha) \Omega_{R'}(\beta)t^R_{a} r 
+ t^{R}_{a}\Omega_{R}(\alpha) \Omega_{R'}(\beta)t^{R'}_{b} r \right. \cr
& \qquad \qquad + r t^{R'}_{b}\Omega_{R}(\alpha) \Omega_{R'}(\beta)t^{R}_{a}
- r t^{R}_{a}\Omega_{R}(\alpha) \Omega_{R'}(\beta)t^{R'}_{b}   \cr
& \qquad \qquad -  t^{R'}_{b}\Omega_{R}(\alpha) \Omega_{R'}(\beta) r t^{R}_{a}
+  t^{R'}_{b} r \Omega_{R}(\alpha) \Omega_{R'}(\beta)t^{R}_{a}   \cr
& \qquad \qquad \left. +  t^{R}_{a}\Omega_{R}(\alpha) \Omega_{R'}(\beta) r t^{R'}_{b}
-  t^{R}_{a} r \Omega_{R}(\alpha) \Omega_{R'}(\beta)t^{R'}_{b}  \right) \cr
& + \frac{1}{8} \kappa^{ab} \kappa^{cd} \tilde s^2  \left( t^{R'}_{b}t^{R'}_{d}\Omega_{R}(\alpha) \Omega_{R'}(\beta)t^R_{c} t^R_{a} 
- 2 t^{R'}_{b}t^{R}_{c}\Omega_{R}(\alpha) \Omega_{R'}(\beta)t^{R'}_{d} t^R_{a} 
\right. \cr & \qquad \qquad \left. 
+ t^{R}_{a}t^{R}_{c}\Omega_{R}(\alpha) \Omega_{R'}(\beta)t^{R'}_{d} t^{R'}_{b}\right)\cr
& +  f^4  \int_0^{2\pi} d\sigma  T^{2\pi,\sigma}_R(\alpha)T^{2\pi,\sigma}_{R'}(\beta)
 i{f^{ac}}_d i{f^{db}}_e 
  \left(
\left( h_1 t^R_{(a}t^R_{b)}t^{R'}_{c} + h_2 t^R_{a}t^{R'}_{(b}t^{R'}_{c)} \right)j^e_z(\sigma) \right.
 \cr
 & \qquad 
+ \left. \left( \bar h_1 t^R_{(a}t^R_{b)}t^{R'}_{c} + \bar h_2 t^R_{a}t^{R'}_{(b}t^{R'}_{c)} \right)j^e_{\bar z}(\sigma) \right) 
T^{\sigma,0}_R(\alpha)T^{\sigma,0}_{R'}(\beta)
\cr & + \mathcal{O}(f^6)
\end{align}
where we introduced $\tilde s$ defined as $s = \tilde s\, \kappa^{ab} t^R_{a}t^{R'}_{b}$.
It is implicit that all integration contours on the right-hand side lie at $\tau=0$.


From the fusion of monodromy matrices \eqref{fusionMM},
we deduce the fusion of transfer matrices by taking a supertrace over the tensor-product of representation $R\otimes R'$.
We obtain:
\begin{align}\label{fusionTrTr}
\lim_{\epsilon \to 0^+} \T_{R}^{(\epsilon)}&(\alpha) \T_{R'}^{(0)}(\beta) =
\T_{R}(\alpha) \T_{R'}(\beta) 
 \cr
& +  f^4  STr \left( \int_0^{2\pi} d\sigma  T^{2\pi,\sigma}_R(\alpha)T^{2\pi,\sigma}_{R'}(\beta)
 i{f^{ac}}_d i{f^{db}}_e \right.
 \left( \left( h_1 t^R_{(a}t^R_{b)}t^{R'}_{c} + h_2 t^R_{a}t^{R'}_{(b}t^{R'}_{c)} \right)j^e_z(\sigma) \right.
  \cr & \qquad 
+\left. \left. \left( \bar h_1 t^R_{(a}t^R_{b)}t^{R'}_{c} + \bar h_2 t^R_{a}t^{R'}_{(b}t^{R'}_{c)} \right)j^e_{\bar z}(\sigma) \right) 
T^{\sigma,0}_R(\alpha)T^{\sigma,0}_{R'}(\beta) \right)
\cr &
 + \mathcal{O}(f^6)
\end{align}
Once again all integration contours on the right-hand side lie at $\tau=0$.
The $r,s$ matrices that were previously associated with the endpoints of the intervals do not appear in the fusion of transfer matrices.


\section{The Hirota equation from the fusion of transfer matrices}\label{proofHirota}

In this section we use the results obtained in section \ref{fusion}, in particular equation \eqref{fusionTrTr}, to give a perturbative derivation of the Hirota equation \eqref{Hirota}.

The classical monodromy matrix is an element of the supergroup $PSl(n|n)$.
Consequently the classical transfer matrix is the character of this group element.
In this section we consider only unitary representations associated to rectangular Young tableaux.
These representations are labeled by two integers $(a,s)$. 
These indices take values in a ``T-hook'' lattice, which precise shape depends on which real form of the supergroup is considered.
For the real form $PSU(p,n-p|n)$ the left-wing of the T-hook has width $p$, the right-wing has width $n-p$ and the vertical strip has width $n$.
More details can be found in \cite{Gromov:2010vb} (see also \cite{Volin:2010xz}).

Characters of $PSl(n|n)$ associated to rectangular Young tableaux satisfy the Jacobi-Trudi formula (see appendix \ref{charIdentities}, in particular equation \eqref{charId(1)PSL}).
This implies for the classical transfer matrices:
\be \T_{a,s}^{clas.}(\alpha)\T_{a,s}^{clas.}(\alpha) = 
 \T_{a,s+1}^{clas.}(\alpha)\T_{a,s-1}^{clas.}(\alpha) +  \T_{a+1,s}^{clas.}(\alpha)\T_{a-1,s}^{clas.}(\alpha) \ee
The previous equation receives quantum corrections that come from the process of fusion of the transfer matrices involved.
Next we will show that the effect of the quantum corrections is to shift the arguments of the transfer matrices. More precisely we will derive the Hirota equation \eqref{Hirota}. The proof relies on character identities for the supergroup $PSl(n|n)$ that can be found in \cite{Kazakov:2007na}. These identities are presented in appendix \ref{charIdentities}.

Let us consider the fusion of the following transfer matrices:
\begin{align}\label{fusion3TT}
 \lim_{\epsilon \to 0^+}
\T^{(\epsilon)}_{a,s}(\alpha)\T^{(0)}_{a,s}(\beta) -  \T^{(\epsilon)}_{a+1,s}(\alpha)\T^{(0)}_{a-1,s}(\beta) - \T^{(\epsilon)}_{a,s-1}(\alpha)\T^{(0)}_{a,s+1}(\beta)
\end{align}
We evaluate this quantity thanks to equation \eqref{fusionTrTr}.
In order to simplify the following equations let us introduce the notation:
\begin{align}\label{defH} \H(x \otimes y) = 
STr_{(a,s)\otimes(a,s)}(x \otimes y) - STr_{(a+1,s)\otimes(a-1,s)}(x \otimes y) - STr_{(a,s-1)\otimes(a,s+1)}(x \otimes y)
\end{align}
In the previous definition $x$ and $y$ are group elements, or the product of group elements with some generators.
The result of the fusion \eqref{fusion3TT} then reads:
\begin{align}\label{fusion3TTstep2}
&\H(\Omega(\alpha)\otimes \Omega(\beta)) 
 +  f^4  \H \left( \int_0^{2\pi} d\sigma  T^{2\pi,\sigma}(\alpha)\otimes T^{2\pi,\sigma}(\beta)
 i{f^{ac}}_d i{f^{db}}_e \right. \cr
 & \qquad \times \left(
\left( h_1 t_{(a}t_{b)}\otimes t_{c} + h_2 t_{a}\otimes t_{(b}t_{c)} \right)j^e_z(\sigma) 
+ \left( \bar h_1 t_{(a}t_{b)}\otimes t_{c} + \bar h_2 t_{a}\otimes t_{(b}t_{c)} \right)j^e_{\bar z}(\sigma) \right) 
\cr & \qquad \times 
 \left.T^{\sigma,0}(\alpha)\otimes T^{\sigma,0}(\beta) \right)
 + \mathcal{O}(f^6)
\end{align}
Now we assume that that $\alpha-\beta = \mathcal{O}(f^2)$.
In that case the coefficients $N_R$ and $N_{R'}$ defined in \eqref{NN'} are of order $f^{-2}$.
Consequently the coefficients $h_1$, $h_2$, $\bar h_1$ and $\bar h_2$ are also of order $f^{-2}$.
Accordingly we focus on the terms of order $f^2$ in \eqref{fusion3TTstep2}%
\footnote{One may ask about the order of magnitude of the subleading terms in \eqref{fusion3TTstep2} contained in the $\mathcal{O}(f^6)$ when $\alpha-\beta = \mathcal{O}(f^2)$. The terms of order $f^{4p+2}$ are expected to vanish since they contribute to the commutator of the transfer matrices only. Under this assumption we expect that the terms of order $f^{4p}$ in \eqref{fusion3TTstep2} become of order $f^{2p}$ in the limit $\alpha-\beta = \mathcal{O}(f^2)$. This is suggested by a sketchy generalization of the computation presented in section \ref{fusionOrder2}.}%
.

To get the leading contribution from the 
second term in \eqref{fusion3TTstep2} we can assume that the transfer matrices are evaluated with the same spectral parameter. Then the character identities \eqref{charId(2)ffJPSL} allows to rewrite \eqref{fusion3TTstep2} as:
\begin{align}\label{fusion3TTstep3}
&\H(\Omega(\alpha)\otimes \Omega(\beta))  +  f^4  \H \left( \int_0^{2\pi} d\sigma  T^{2\pi,\sigma}(\alpha)\otimes T^{2\pi,\sigma}(\alpha) 
 \left( \left( h_1  + h_2  \right)j^e_z(\sigma) 
+ \left( \bar h_1  + \bar h_2 \right)j^e_{\bar z}(\sigma) \right) 
\right. \cr
 & \qquad \times \left. 
(t_e \otimes 1 - 1 \otimes t_e)
T^{\sigma,0}(\alpha)\otimes T^{\sigma,0}(\alpha) \right)
 + \mathcal{O}(f^4)
\end{align}
We notice that the coefficients simplify:
\begin{align}
& h_1+h_2 = \pi^2 \left( \frac{2}{1+\alpha}\right)^2 \frac{4}{\alpha-\beta} \frac{(1+\alpha-2\eta)^4}{(1-\alpha^2)^2} +\mathcal{O}(f^0)\cr
& \bar h_1+\bar h_2 = -\pi^2 \left( \frac{2}{1-\alpha}\right)^2 \frac{4}{\alpha-\beta} \frac{(1+\alpha-2\eta)^4}{(1-\alpha^2)^2} +\mathcal{O}(f^0)
\end{align}
This remarkably implies that the combination of currents appearing in \eqref{fusion3TTstep3} is proportional to the derivative of the flat connection with respect to the spectral parameter:
\be (h_1+h_2) j^e_z + (\bar h_1 + \bar h_2) j^e_{\bar z} 
= - \frac{8\pi^2}{\alpha-\beta} \frac{(1+\alpha-2\eta)^4}{(1-\alpha^2)^2} \p_\alpha A^e(\alpha)+\mathcal{O}(f^0) \ee
So \eqref{fusion3TTstep3} can be rewritten as:
\begin{align}\label{fusion3TTstep4}
&\H(\Omega(\alpha)\otimes \Omega(\beta)) 
-  f^4  \frac{8\pi^2}{\alpha-\beta} \frac{(1+\alpha-2\eta)^4}{(1-\alpha^2)^2}
\left(\H(\p_\alpha \Omega(\alpha)\otimes \Omega(\alpha))-\H(\Omega(\alpha)\otimes \p_\alpha \Omega(\alpha))
\right)
 + \mathcal{O}(f^4)
\end{align}
Remember that $\H(M(\alpha)\otimes M(\alpha))=0$, so both terms in the previous equation are of order $f^2$.
Let us now introduce:
\begin{align} & \alpha_\pm = \alpha \mp 2 \pi f^2 \frac{(1+\alpha-2\eta)^2}{1-\alpha^2} 
\end{align}
We consider equation \eqref{fusion3TTstep4} with $\alpha=\alpha_+$ and $\beta=\alpha_-$. Then the order-$f^2$ terms cancel. Thus we obtain the Hirota equation, up to corrections of order $f^4$:
\begin{align}\label{almostHirota} &\lim_{\epsilon \to 0^+}
\T_{a,s}^{(\epsilon)}(\alpha_+)\T_{a,s}^{(0)}(\alpha_-) -  \T_{a+1,s}^{(\epsilon)}(\alpha_+)\T_{a-1,s}^{(0)}(\alpha_-) - \T_{a,s-1}^{(\epsilon)}(\alpha_+)\T_{a,s+1}^{(0)}(\alpha_-) = 0 + \mathcal{O}(f^4)
\end{align}
We can perform a change of variables to write the Hirota equation in its usual form. We look for $u$ such that:
\be \alpha(u+1) = \alpha(u) - 2\pi f^2 \frac{(1+\alpha-2\eta)^2}{1-\alpha^2} \ee
A straightforward integration gives:
\be  u = \frac{1}{2\pi f^2} \left( \alpha + 2 (2\eta-1) \log (\alpha+1-2\eta) + \frac{1-(1-2\eta)^2}{\alpha +1 -2\eta}\right) \ee
Notice the simplification in the case $\eta=\frac{1}{2}$, namely when the coefficient of the Wess-Zumino term in the action \eqref{action} is set to zero:
\be\label{ZukVar}  u = \frac{1}{2\pi f^2} \left(\alpha + \frac{1}{\alpha} \right) \ee
In terms of the variable $u$ equation \eqref{almostHirota} reads:
\begin{align} &\lim_{\epsilon \to 0^+}
 \T_{a,s}^{(\epsilon)}(u+1)\T_{a,s}^{(0)}(u-1) -  \T_{a+1,s}^{(\epsilon)}(u+1)\T_{a-1,s}^{(0)}(u-1) - \T_{a,s-1}^{(\epsilon)}(u+1)\T_{a,s+1}^{(0)}(u-1) \cr
& \qquad = 0 + \mathcal{O}(f^4)
\end{align}
This is the Hirota equation \eqref{Hirota}.

\paragraph{About the spectrum.}
The Hirota equation typically admits an infinite number of solutions with different analytic properties.
It is believed that each states in the spectrum of the theory can be associated to one of these solutions (see for instance \cite{Gromov:2008gj}).
The energy of a state can be obtained from the corresponding solution.
For instance in the $AdS_5$/$CFT_4$ case \cite{Gromov:2009tv} the energy is given by:
\be\label{Energy?} E = \sum_j \epsilon^{(p)}_1(u_{4,j}) + \sum_{a=1}^\infty \int_{-\infty}^\infty \frac{du}{2\pi i} \p_u \epsilon^{(m)}_a \log(1+Y^{(m)}_{a,0}) \ee
The $Y$-functions are related to the $\T$-functions as:
\be Y_{a,s} = \frac{\T_{a,s+1}\T_{a,s-1}}{\T_{a+1,s}\T_{a-1,s}} \ee
The functions $\epsilon_a(u)$ are given by:
\be \epsilon_a(u) = a + \frac{2ig}{\alpha(u+a)}- \frac{2ig}{\alpha(u-a)} \ee
where $\alpha(u)$ is obtained by inverting a relation similar to \eqref{ZukVar}:
\be \frac{u}{g} = \alpha + \frac{1}{\alpha} \ee
The quantities $u_{4,j}$ satisfy $Y_{1,0}(u_{4,j})=-1$.
Eventually the labels $(p)$ and $(m)$ in \eqref{Energy?} means that the functions have to be evaluated in the so-called physical or mirror kinematics.

Formula \eqref{Energy?} follows from the Thermodynamic Bethe Ansatz approach. 
Most likely a very similar formula is also valid in the models we are considering.
Probably such a formula can be derived using the Thermodynamic Bethe Ansatz techniques.
However it would be more satisfactory to have a first-principle derivation of such a formula, which would be closer in spirit to the approach of this paper.
We hope to report on this issue in the future.


\section{Conclusion}

Let us briefly summarize the results derived in this paper.
First we studied the divergences that appear in line operators up to second order in perturbation theory.
We showed that these divergences can be canceled with a simple wave-function renormalization of these operators. 
Moreover the transfer matrix is free of any divergences.
In a second time we computed the fusion of line operators up to second order in perturbation theory.
The result is given in equation \eqref{fusionf4}, from which we deduced the fusion of monodromy and transfer matrices.
Most of the computations presented in section \ref{fusion} can be translated straightforwardly for any theory with a flat connection that realizes a $r,s$ Maillet system, assuming the self-OPE of the flat connection is known up to a sufficient order in perturbation theory.
Eventually we used these results to prove that the transfer matrix satisfies the Hirota equation \eqref{Hirota}.
In particular we showed that the shift of the spectral parameter in the Hirota equation is a quantum effect resulting from the simultaneous collision of three connections or more.

The approach advocated in this paper can be useful to get a better understanding of the quantum integrable structure of relativistic sigma-models.
The path we followed to derive the Hirota equation is much more direct than the usual Thermodynamic Bethe Ansatz method.
In particular we did not assume quantum integrability of the model, and there is obviously no need for the string hypothesis. 
As such our method can be used to check of the validity of the assumptions needed in the Thermodynamic Bethe Ansatz approach.
Further work is needed to reach a solution for the spectrum which is completely independent of the Thermodynamic Bethe Ansatz approach.
In particular it would be very interesting to get a first-principles derivation of a formula that gives the energy in terms of the $\T$-functions.

It should be rather straightforward to extend the results we derived to other integrable relativistic sigma-models.
The only ingredient needed is essentially the quantum OPE between two flat connections.
An obvious candidate is the pure-spinor string on $AdS_5\times S^5$. This theory seems to realize a $r,s$ Maillet system \cite{Mikhailov:2007eg}\cite{Magro:2008dv} and the transfer matrix is free of divergences, at least up to first order in perturbation theory \cite{Mikhailov:2007mr}. So most of the results of this paper can be translated directly for the pure spinor string on $AdS_5\times S^5$.
Another interesting example is the hybrid string on $AdS_3\times S^3 \times T^4$ \cite{Berkovits:1999im}. This theory is essentially the sigma-model on $PSU(1,1|2)$ coupled to ghosts. So to generalize our results to this case one has to dress-up the current-current OPEs \eqref{jjOPEs} with the hybrid ghosts. This would be an interesting step to study the $AdS_3$/$CFT_2$ correspondence with integrability methods, but also more generally to understand the quantization of string theory in RR backgrounds.

\section*{Acknowledgments}

The author has a large debt towards Pedro Vieira for several crucial discussions concerning this project.
The author would also like to thank Gleb Arutyunov, Fran\c cois Delduc, Marc Magro, Joerg Teschner, Jan Troost, Benoit Vicedo and an anonymous referee for useful discussions and correspondence.
The author is a Postdoctoral researcher of FWO-Vlaanderen.
This research is supported in part by the Belgian Federal Science Policy Office through the Interuniversity Attraction Pole IAP VI/11 and by FWO-Vlaanderen through project G011410N.

\begin{appendix}


\section{Hamiltonian formalism and Maillet $r,s$ systems}\label{Maillet}

In this appendix we briefly review the computation of the Poisson-bracket of transition matrices in a family of classically integrable models using the Hamiltonian formalism.
More details can be found for instance in \cite{Dorey:2006mx}. 
We also explain the link between the Hamiltonian approach and the first-order computations of section \ref{fusion}.

We start with a Lax connection $A(\alpha;\sigma)$:
\be dA(\alpha;\sigma) + A(\alpha;\sigma)\wedge A(\alpha;\sigma) = 0 \ee
The monodromy matrix involves the integral of the space component of the Lax connection, that we also denote by $A(\alpha)$. We assume that the Poisson bracket between two space components of $A(\alpha)$, taken in arbitrary and possibly different representations, is of the form:
\begin{align}\label{PoissonJJ}
\{ A(\alpha;\sigma),A(\beta;\sigma') \} =&
(r(\sigma,\alpha,\beta) + s(\sigma,\alpha,\beta) - r(\sigma',\alpha,\beta) + s(\sigma',\alpha,\beta)) \delta'(\sigma-\sigma') \cr
&+ [A(\alpha;\sigma)\otimes 1 + 1 \otimes A(\beta;\sigma'),r(\sigma,\alpha,\beta)] \delta(\sigma-\sigma') \cr
&+ [A(\alpha;\sigma)\otimes 1 - 1 \otimes A(\beta;\sigma'),s(\sigma,\alpha,\beta)] \delta(\sigma-\sigma') 
\end{align}
The previous equation defines a $r,s$ Maillet system.
We consider the transition matrix:
\be T^{b,a}(\alpha) = P \exp \left( -\int_{a}^{b} d \sigma A(\alpha;\sigma) \right) \ee
The Poisson bracket of transition matrices evaluates to:
\begin{align}\label{poissonTT}
\{ T^{b,a}(\alpha),& T^{d,c}(\beta)\}  
= \epsilon(d-c) \chi(b;c,d)(1\otimes T^{d,b}(\beta)(r(b,\alpha,\beta)+s(b,\alpha,\beta))T^{b,a}(\alpha)\otimes T^{b,c}(\beta))\cr
& - \epsilon(d-c) \chi(a;c,d)(T^{b,a}(\alpha)\otimes T^{d,a}(\beta)(r(a,\alpha,\beta)+s(a,\alpha,\beta))1\otimes T^{a,c}(\beta))\cr
& + \epsilon(b-a) \chi(d;a,b)(T^{b,d}(\alpha)\otimes 1(r(d,\alpha,\beta)-s(d,\alpha,\beta))T^{c,a}(\alpha)\otimes T^{d,c}(\beta))\cr
& - \epsilon(b-a) \chi(c;a,b)(T^{b,c}(\alpha)\otimes T^{d,c}(\beta)(r(c,\alpha,\beta)-s(c,\alpha,\beta))T^{c,a}(\alpha)\otimes 1) 
\end{align}
where $\epsilon(\sigma)$ is the sign function, and $\chi(\sigma;a,b)$ is the characteristic function of the interval $[a,b]$.
The previous formula is ambiguous when the endpoints of the integration contours are not all distinct.
The Maillet prescription \cite{Maillet:1985ek} is to perform a symmetric point-splitting of the coinciding endpoints to resolve the ambiguity.

The Poisson bracket does not satisfy the Jacobi identity.
To solve this problem Maillet suggested to define the Poisson bracket in a weak way: the definition of the bracket depends on how many brackets are nested. The correct prescription is then to average in a symmetric way over all possible point-splittings.

Now let us consider the monodromy matrix. Space is compactified: $\sigma=\sigma+2\pi$. The monodromy matrix is:
\be \Omega(\alpha) = T^{2\pi,0}(\alpha) \ee
We want to compute the Poisson bracket of two monodromy matrices. We work on the universal cover of the cylinder. We have to sum over all possible lift of the two integration contours to the plane that have a non-trivial overlap:
\begin{align} \{\Omega(\alpha), \Omega(\beta) \} 
&= \{ T^{2\pi,0}(\alpha),T^{2\pi,0}(\beta)\} 
+ \{ T^{2\pi,0}(\alpha),T^{0,-2\pi}(\beta)\} 
+ \{ T^{2\pi,0}(\alpha),T^{4\pi,2\pi}(\beta)\} \end{align}
We obtain:
\begin{align} \{\Omega(\alpha), \Omega(\beta) \} 
=& [r(0,\alpha,\beta),\Omega(\alpha)\otimes  \Omega(\beta) ] 
+ (\Omega(\alpha)\otimes 1) s(0,\alpha,\beta) (1\otimes  \Omega(\beta) ) \cr
&
- (1\otimes  \Omega(\beta) ) s(0,\alpha,\beta) (\Omega(\alpha)\otimes 1) \end{align}
We deduce that the Poisson bracket of transfer matrices is zero:
\begin{align} \{Tr \Omega(\alpha), Tr \Omega(\beta) \} 
=0 \end{align}

\subsection{Dictionary OPE/Commutators}\label{OPE[]}

We define the commutator of equal-time operators as:
\be [A(\sigma),B(0)] = \lim_{\epsilon \to 0^+}  \left( A(\sigma+i \epsilon)B(0) - B(i \epsilon)A(\sigma) \right) \ee
We extract an operative dictionary between OPEs and commutators.
Let us consider for instance the following OPE:
\be\label{ABOPE} A(x) B(0) = \frac{C}{x^2} + \frac{D}{\bar x^2} + E \delta^{(2)}(x) + \frac{F(0)}{x} + \frac{G(0) \bar x}{x^2} + \frac{H(0)}{\bar x} + \frac{I(0) x}{\bar x^2} + ... \ee
It is translated to:
\be\label{ABcom} \frac{1}{2\pi i} [ A(\sigma), B(0) ] = C \delta'(\sigma) - D \delta'(\sigma) - F(0) \delta(\sigma) - G(0)\delta(\sigma) + H(0)\delta(\sigma) + I(0)\delta(\sigma) \ee
This result is straightforwardly generalized to the case where the OPE contains poles of order greater than two.
Notice that the OPE contains more information than the commutator.
This dictionary shows that the first-order computation of fusion presented in section \ref{fusionOrder1} is equivalent to the computation of the Poisson bracket of line operators in the Hamiltonian formalism. 

\paragraph{About the sub-leading singular terms.}
The OPE \eqref{ABOPE} generically contains additional singular terms that come with operators of conformal dimension greater or equal to the sum of the conformal dimensions of the operators on the left-hand side.
We loosely call such terms ``sub-leading singular terms''.
In section \ref{rsMatrices} we argued using locality and dimensional analysis that the sub-leading singular terms in the current-current OPE do not contribute to the commutator of two connections. 
Here we will give a more elementary proof of this statement.
Consider a sub-leading singularity in the OPE \eqref{ABOPE}, for instance:
\be A(x) B(0) = ... + O(0) \frac{x^n}{\bar x^{\bar n}} +... \ee
with $n \ge \bar n$.
The contribution of this term to the commutator $[ A(\sigma), B(0) ]$ is proportional to:
\be \lim_{\epsilon \to 0^+} \left(
 \frac{(\sigma+i \epsilon)^n}{(\sigma-i \epsilon)^{\bar n}} - \frac{(\sigma-i \epsilon)^n}{(\sigma+i \epsilon)^{\bar n}} 
 \right)
 =0 \ee
A logarithmic dependance on the coordinates may also appear in the sub-leading singular terms. It does not modify the previous result:
\be \lim_{\epsilon \to 0^+} \left(
 \frac{(\sigma+i \epsilon)^n}{(\sigma-i \epsilon)^{\bar n}}\log^m(\sigma+i \epsilon)\log^{\bar m}(\sigma-i \epsilon) 
 - \frac{(\sigma-i \epsilon)^n}{(\sigma+i \epsilon)^{\bar n}} \log^m(\sigma-i \epsilon)\log^{\bar m}(\sigma+i \epsilon)
 \right)
 =0 \ee
Here we still assumed that $n \ge \bar n$.
We conclude that none of the sub-leading singular terms in the OPE \eqref{ABOPE} do contribute to the commutator \eqref{ABcom}.


\section{Divergences in the transfer matrix in a generic WZW model}\label{divWZW}

In this appendix we consider a two-dimensional sigma-model defined by the action \eqref{action} on a generic Lie-group $G$.
We do not assume the vanishing of the dual Coxeter number of the group.
Such a model is conformal invariant only at the WZW points, i.e. for $\eta=1$ or $\eta=0$. For definiteness we pick the first choice.
For other values of $\eta$ the current-current OPEs \eqref{jjOPEs} receive corrections.

At the WZW points only one components of the current is non-zero. It is usual to use the notation $J$ for the holomorphic current. The OPEs \eqref{jjOPEs} simplify to%
\footnote{Our normalization for the current differs from most of the WZW literature by a factor of $f^2=\frac{1}{k}$.}%
:
\be J^a(z) J^b(0) = f^2 \frac{\kappa^{ab}}{z^2} + f^2 i \frac{{f^{ab}}_c J^c(0)}{z} + ... \ee
The one-parameter family of flat connections is simply:
\be A(\lambda;z) = \lambda J(z) \ee
where $\lambda$ is the spectral parameter. The transfer matrix read:
\be \T(\lambda) = \exp \left( - \oint A(\lambda) \right) \ee
It is convenient to expand the transfer matrix as:
\be \T(\lambda) = \sum_{N=0}^\infty (-1)^N T_N(\lambda) \ee

In this appendix we study the divergences that appear in the transfer matrix in a generic WZW model. This problem was previously considered in \cite{Bachas:2004sy} and \cite{Alekseev:2007in}. An important result of these papers is that the transfer matrix can be consistently quantized for a particular value of the spectral parameter. In this special case the transfer matrix is the trace of the monodromy of the solution to the equations of motion. In our conventions this special value of the spectral parameter is $\lambda=1$. Our goal is to recover this result, up to second order in perturbation theory, using the regularization prescription introduced in section \ref{regularization}. Actually we essentially follow the computations presented in section \ref{renormalization}, only keeping track of the terms proportional to the dual Coxeter number. We denote by $\check h$ the dual Coxeter number of the Lie-group $G$.

First we study the divergences that appear at first order, i.e. after we perform one single OPE.

\paragraph{First-order pole.}
The first-order pole produces a divergence in the OPE between two adjacent connections in $\T_N$ (see Figure \ref{div1}(b)). This leads to divergent corrections to the classical transfer matrix of the form:
\be -N f^2 \lambda \check h \log \epsilon \T_{N} \ee

\paragraph{Second-order pole.}
The second-order pole produces a divergence in the OPE between two adjacent (see Figure \ref{div1}(c) and Figure \ref{div2M}(b)) or next-to adjacent (see Figure \ref{div1}(d)) connections in $\T_N$. Divergences proportional to the quadratic Casimir vanish. We are left with divergent corrections of the form:
\be N f^2 \lambda^2 \check h \log \epsilon \T_{N} \ee

\paragraph{Renormalization of the transfer matrix at first-order.}
The previous divergences can be canceled by the following renormalization of the transfer matrix:
\be \T(\lambda) \to \T(\lambda_R) \ee
with:
\be \lambda_R = \lambda - f^2  \check h \log \epsilon \lambda(\lambda-1) \ee
Notice that the transfer matrix is not renormalized when $\lambda=1$.

Next we study the divergences that appear at second order.

\paragraph{OPEs between distinct pairs of connections.}
From the previous results we deduce straightforwardly the second-order corrections resulting from two OPEs that involve distinct pairs of connections:
\be (f^2\check h \log \epsilon)^2 \lambda(\lambda-1) N ((N+2)\lambda-(N+1)) \T_{N} \ee
They obviously vanish when $\lambda=1$.

\paragraph{Triple OPEs.}
Now we consider the divergences resulting from triple OPEs. We will be slightly schematic here. In particular we do not pretend to perform an exhaustive analysis of all possible terms. We will rather identify a generic pattern of cancellations between the divergent terms when $\lambda=1$. The divergent terms we obtain come with the following operator:
\begin{align} & \tilde \T_{N;a,b,c}^{i,j} = \oint_{\sigma_1>...>\sigma_N}
A(\lambda;\sigma_1)...A(\lambda;\sigma_{i-1}) j_a(\sigma_i)t_b A(\lambda;\sigma_{i+1})...A(\lambda;\sigma_{j-1}) t_c \cr & \quad \times A(\lambda;\sigma_{j+1})...A(\lambda;\sigma_{N})P.V.\frac{1}{\sigma_i-\sigma_j} \end{align}
Let us first discuss the divergences coming from the first-order pole in the second OPE. For a divergence to appear at least two of the three connections involved have to be adjacent. We obtain divergent terms of the form:
\be\label{div2L3} \frac{3}{2} \lambda^3 f^4 \check h \log \epsilon i f^{abc} \T_{N;a,b,c}^{i,j} \ee
The case where the connections involved in the first (resp. second) OPE are adjacent contribute to one (resp. $\frac{1}{2}$) to the factor $\frac{3}{2}$.
Next we discuss the divergences coming from the second-order pole in the second OPE. They lead to the divergent terms of the form: 
\be\label{div2L4} -\frac{3}{2} \lambda^4 f^4 \check h \log \epsilon i f^{abc} \T_{N;a,b,c}^{i,j} \ee
The case where the connections involved in the first (resp. second) OPE are adjacent contribute to one (resp. $\frac{1}{2}$) to the factor $\frac{3}{2}$. 
We observe that the two terms \eqref{div2L3} and \eqref{div2L4} cancel against each other when $\lambda=1$. 
For generic $\lambda$ it is not obvious that these divergent terms can be canceled by a renormalization of spectral parameter combined with a wave-function renormalization of the transfer matrix.

\paragraph{Upshot.}
In this appendix we have given some evidence that the divergent terms that appear in the transfer matrix of a generic WZW model do vanish for a special value of the spectral parameter, in agreement with the results of \cite{Bachas:2004sy} and \cite{Alekseev:2007in}. 
The divergences coming from the double pole generically cancel against the divergences coming from the simple pole for $\lambda=1$.
This gives some support in favor of the generic validity of the techniques used in section \ref{renormalization}, in particular concerning the choice of points at which the operators are evaluated in the current-current OPEs \eqref{jjOPEs}.


\section{Another regularization of the line operators}\label{altRegScheme}

In this appendix we discuss another possible regularization of the transition matrices.
The UV divergences are regularized by constraining the distance between integrated operators to be greater than a cut-off $\epsilon'$:
\be\label{altScheme} \int_{b>z_1>z_2>a}dz_1 dz_2 O(z_1) O(z_2) \to \int_{a}^b dz_1 \int_{a}^{z_1-\epsilon'} dz_2 O(z_1) O(z_2)\ee
This regularization prescription is much sharper than the one introduced in section \ref{regularization}.
Using this regularization scheme, it was shown in  \cite{Mikhailov:2007mr} that the transfer matrix for the pure-spinor string on $AdS_5 \times S^5$ has no logarithmic divergences, but has linear divergences, at first order in perturbation theory. We will show that the same conclusion applies for the models studied in the present paper.

We consider the divergences appearing in the following operator:
\be \int_{b>\sigma_1>...>\sigma_N>a}d\sigma_1...d\sigma_N A^{a_1}(\alpha;\sigma_1)...A^{a_N}(\alpha;\sigma_N) t_{a_1}...t_{a_N} \ee
The OPE between connections is:
\begin{align} A^a(\alpha;\sigma) A^b(\alpha;\sigma') & = p_2 \kappa^{ab} \frac{1}{(\sigma-\sigma')^2} 
 + p_1 {f^{ab}}_c (j_z^c(\sigma)+j_z^c(\sigma')) \frac{1}{\sigma-\sigma'} \cr
 & + \bar p_1 {f^{ab}}_c (j_{\bar z}^c(\sigma)+j_{\bar z}^c(\sigma')) \frac{1}{\sigma-\sigma'} + ... \end{align}

\paragraph{Divergences from first-order poles}
The first order pole in the OPE between adjacent connections leads to logarithmic divergences. All these divergences vanish since ${f^{ab}}_c t_a t_b = 0$.

\paragraph{Linear divergences from second-order poles}
The second-order poles in the OPE between adjacent connections lead to linear divergences (and also to logarithmic divergences that we discuss later).
These are proportional to:
\be p_2 \kappa^{ab}t_a t_b \frac{1}{\epsilon'} = p_2 C^{(2)} \frac{1}{\epsilon'}\ee
These divergences can be canceled by a scalar wave-function renormalization of the transition matrices:
\be T_R^{a,b}(\alpha) = Z_R P \exp \left( - \int_a^b A_R(\alpha) \right) \ee
where $Z_R$ is equal to:
\be Z_R = \exp \left( -\frac{b-a}{\epsilon'} p_2 C^{(2)}_R  \right) \ee

\paragraph{Logarithmic divergences from second-order poles}
The second-order poles in the OPE between connections separated by one or zero connections lead to logarithmic divergences.
These divergences are exactly the same than the one we got in section \ref{divOrder1} using the ``soft'' regularization scheme.
In particular they add up to zero.

\paragraph{Upshot}
The regularization scheme \eqref{altScheme} gives rise to the same logarithmic divergences than the smooth regularization scheme introduced in section \ref{regularization}. However new linear divergences appear. Their cancellation require a renormalization of the line operators. In particular with the regularization scheme \eqref{altScheme} the transfer matrix has to be renormalized with a representation-dependent factor.


\section{Technical details about the fusion of line operators}

In this appendix we give additional details about the computations described in section \ref{fusion}.


\subsection{First-order computations}\label{CompFusionO1}
Here we describe the computation that leads to equation \eqref{fusionf2}

\paragraph{A warm up.}
To get some intuition about the generic pattern of simplifications it is useful to study a simple case first.
Let us evaluate the terms in \eqref{expansionInts} with one integral only.
At order $f^0$ these terms are simply $T^{b,a}_{R,1} + T^{d,c}_{R',1}$. At order $f^2$ we have three contributions: 
\begin{itemize}
	\item One from $T^{b+i\epsilon,a+i\epsilon}_{R,1} T^{d,c}_{R',1}$, using the first-order singularity in the OPE \eqref{OPEAA'delta}.
	\item One from $-T^{b+i\epsilon,a+i\epsilon}_{R,2} T^{d,c}_{R',1}$, using the second-order singularity in the OPE \eqref{OPEAA'delta}.
	\item One from $-T^{b+i\epsilon,a+i\epsilon}_{R,1} T^{d,c}_{R',2}$, using the second-order singularity in the OPE \eqref{OPEAA'delta}.
\end{itemize}
We will now compute these three contributions.
We start with the simplest case where quantum corrections can appear, namely the OPE between the two integrated connections in $T^{b+i\epsilon,a+i\epsilon}_{R,1} T^{d,c}_{R',1}$.
We use the OPE \eqref{OPEAA'delta}.
We obtain:
\begin{align} 
\int_{[a,b]\cap[c,d]} d\sigma \left( \left[ A_{R}(\alpha;\sigma),\frac{r+s}{2} \right] +  \left[A_{R'}(\beta;\sigma),\frac{r-s}{2}\right] \right)
\end{align}
Let us now consider the contribution from $-T^{b+i\epsilon,a+i\epsilon}_{R,2} T^{d,c}_{R',1}$.
We perform one OPE between two connections and isolate the contribution of the second-order singularity:
\begin{align}
- \int \int_{b>\sigma_1>\sigma_2>a} d\sigma_1  d\sigma_2 \int_c^d d\sigma' & \left(
s \delta'_\epsilon(\sigma_1-\sigma')  A_R(\alpha,\sigma_2) 
+ A_R(\alpha,\sigma_1) s \delta'_\epsilon(\sigma_2-\sigma')
\right)
\end{align}
Then we proceed with the integrations.
We obtain:
\begin{align}
& -\int_{[a,b]\cap [c,d]} d\sigma [A_R(\alpha,\sigma),s] 
- \chi(b;c,d) \int_a^b d\sigma s A_R(\alpha,\sigma)
+ \chi(a;c,d) \int_a^b d\sigma A_R(\alpha,\sigma) s
\end{align}
where $\chi(x;y,z)$ is the characteristic function of the interval $[y,z]$. 
Subtleties arising when the endpoints of the integration path of the transition matrices coincide are discussed in section \ref{singularFusion}.
Eventually the third term that contributes comes from  $-T^{b+i\epsilon,a+i\epsilon}_{R,1} T^{d,c}_{R',2}$.
Following the same steps we obtain:
\begin{align}
& \int_{[a,b]\cap [c,d]} d\sigma [A_{R'}(\beta,\sigma),s] 
+ \chi(d;a,b) \int_c^d d\sigma s A_{R'}(\beta,\sigma)
- \chi(c;a,b) \int_c^d A_{R'}(\beta,\sigma) s
\end{align}

Gathering the three terms we just computed, we get:
\begin{align} 
& \int_{[a,b]\cap[c,d]} d\sigma \left(\left[A_{R}(\alpha;\sigma),\frac{r-s}{2}\right]+\left[A_{R'}(\beta;\sigma),\frac{r+s}{2}\right]\right) \cr
& - \chi(b;c,d) \int_a^b d\sigma  s A_{R}(\alpha;\sigma) + \chi(a;c,d)\int_a^b d\sigma A_{R}(\alpha;\sigma) s  \cr
&  + \chi(d;a,b) \int_c^d d\sigma s A_{R'}(\beta;\sigma) - \chi(c;a,b)\int_c^d d\sigma A_{R'}(\beta;\sigma) s 
\end{align}
We can rearrange the previous expression using the formulas \eqref{cap=3chi} in appendix \ref{intCap}.
For instance we rewrite:
\begin{align}
& \int_{[a,b]\cap[c,d]}d\sigma A_{R}(\alpha;\sigma)\frac{r-s}{2} + \chi(a;c,d)\int_a^b d\sigma A_{R}(\alpha;\sigma) s \cr
&= \chi(a;c,d) \int_a^b d\sigma A_{R}(\alpha;\sigma)\frac{r+s}{2}  + \chi(c;a,b) \int_c^b d\sigma A_{R}(\alpha;\sigma)\frac{r-s}{2} \cr
& \quad - \chi(d;a,b) \int_d^b d\sigma A_{R}(\alpha;\sigma)\frac{r-s}{2} 
\end{align}
At the end we obtain:
\begin{align}\label{TT->A+A'}  
- & \left( \chi(b;c,d) \left( \int_b^d A_{R'}(\beta) \frac{r+s}{2} + \frac{r+s}{2} \int_a^b  A_{R}(\alpha) + \frac{r+s}{2} \int_c^b A_{R'}(\beta) \right) \right. \cr
& - \chi(a;c,d) \left( \int_a^b A_{R}(\alpha) \frac{r+s}{2} + \int_a^d A_{R'}(\beta)\frac{r+s}{2} + \frac{r+s}{2} \int_c^a A_{R'}(\beta) \right) \cr
& + \chi(d;a,b) \left( \int_d^b A_{R}(\alpha) \frac{r-s}{2} + \frac{r-s}{2} \int_a^d A_{R}(\alpha) + \frac{r-s}{2} \int_c^d A_{R'}(\beta) \right) \cr
& \left. - \chi(c;a,b) \left( \int_c^b A_{R}(\alpha) \frac{r-s}{2} + \int_c^d A_{R'}(\beta) \frac{r-s}{2} + \frac{r-s}{2} \int_a^c A_{R}(\alpha) \right)\right)  \end{align}

\paragraph{The generic case.}
Now we explain how to add the three terms \eqref{fusionTTO1Term1},\eqref{fusionTTO1Term2} and \eqref{fusionTTO1Term3} to obtain equation \eqref{fusionf2nn'}.
First we consider \eqref{fusionTTO1Term1}. We expand the commutator. Then we distinguish cases depending on which connections ($A$ or $A'$) are the closest to the constant matrix $r-s$ along the integration path:
\begin{align}
&(-)^{n+n'+1} \sum_{i=0}^{n} \sum_{i'=1}^{n'-1} \int_{[a,b]\cap[c,d]} d\sigma_1 \int_c^{\sigma_1} d \sigma_2
 \left\lfloor \int_{\sigma_1}^b A \right\rceil^i   \left\lfloor \int_{\sigma_1}^d A' \right\rceil^{i'-1} \cr
& \qquad \qquad \times A_{R'}(\beta;\sigma_1)\frac{r-s}{2}A_{R'}(\beta;\sigma_2)
  \left\lfloor \int_a^{\sigma_2} A \right\rceil^{n-i}   \left\lfloor \int_c^{\sigma_2} A' \right\rceil^{n'-i'-1} \cr
&+(-)^{n+n'+1} \sum_{i=0}^{n-1} \sum_{i'=1}^{n'} \int_{[a,b]\cap[c,d]} d\sigma_1 \int_a^{\sigma_1} d \sigma_2
 \left\lfloor \int_{\sigma_1}^b A \right\rceil^i   \left\lfloor \int_{\sigma_1}^d A' \right\rceil^{i'-1} \cr
& \qquad \qquad \times A_{R'}(\beta;\sigma_1)\frac{r-s}{2}A_{R}(\alpha;\sigma_2)
  \left\lfloor \int_a^{\sigma_2} A \right\rceil^{n-i-1}   \left\lfloor \int_c^{\sigma_2} A' \right\rceil^{n'-i'} \cr
&+(-)^{n+n'+1}  \int_{[a,b]\cap[c,d]} d\sigma
 \left\lfloor \int_{\sigma}^b A \right\rceil^n   \left\lfloor \int_{\sigma}^d A' \right\rceil^{n'-1} 
  A_{R'}(\beta;\sigma)\frac{r-s}{2} \cr
&-(-)^{n+n'+1} \sum_{i=0}^{n} \sum_{i'=1}^{n'-1} \int_{[a,b]\cap[c,d]} d\sigma_2 \int_{\sigma_2}^d d \sigma_1
 \left\lfloor \int_{\sigma_1}^b A \right\rceil^i   \left\lfloor \int_{\sigma_1}^d A' \right\rceil^{i'-1} \cr
& \qquad \qquad \times A_{R'}(\beta;\sigma_1)\frac{r-s}{2}A_{R'}(\beta;\sigma_2)
  \left\lfloor \int_a^{\sigma_2} A \right\rceil^{n-i}   \left\lfloor \int_c^{\sigma_2} A' \right\rceil^{n'-i'-1} \cr
&-(-)^{n+n'+1} \sum_{i=1}^{n} \sum_{i'=0}^{n'-1} \int_{[a,b]\cap[c,d]} d\sigma_2 \int_{\sigma_2}^b d \sigma_1
 \left\lfloor \int_{\sigma_1}^b A \right\rceil^{i-1}   \left\lfloor \int_{\sigma_1}^d A' \right\rceil^{i'} \cr
& \qquad \qquad \times A_{R}(\alpha;\sigma_1)\frac{r-s}{2}A_{R'}(\beta;\sigma_2)
  \left\lfloor \int_a^{\sigma_2} A \right\rceil^{n-i}   \left\lfloor \int_c^{\sigma_2} A' \right\rceil^{n'-i'-1} \cr
&-(-)^{n+n'+1} \int_{[a,b]\cap[c,d]} d\sigma 
 \frac{r-s}{2}A_{R'}(\beta;\sigma)
  \left\lfloor \int_a^{\sigma} A \right\rceil^{n}   \left\lfloor \int_c^{\sigma} A' \right\rceil^{n'-1} 
\end{align}

Similarly we rewrite the contribution \eqref{fusionTTO1Term2} as:
\begin{align}
&(-)^{n+n'+1} \sum_{i=1}^{n-1} \sum_{i'=0}^{n'} \int_{[a,b]\cap[c,d]} d\sigma_1 \int_a^{\sigma_1} d \sigma_2
 \left\lfloor \int_{\sigma_1}^b A \right\rceil^{i-1}   \left\lfloor \int_{\sigma_1}^d A' \right\rceil^{i'} \cr
& \qquad \qquad \times A_{R}(\alpha;\sigma_1)\frac{r+s}{2}A_{R}(\alpha;\sigma_2)
  \left\lfloor \int_a^{\sigma_2} A \right\rceil^{n-i-1}   \left\lfloor \int_c^{\sigma_2} A' \right\rceil^{n'-i'} \cr
&+(-)^{n+n'+1} \sum_{i=1}^{n} \sum_{i'=0}^{n'-1} \int_{[a,b]\cap[c,d]} d\sigma_1 \int_c^{\sigma_1} d \sigma_2
 \left\lfloor \int_{\sigma_1}^b A \right\rceil^{i-1}   \left\lfloor \int_{\sigma_1}^d A' \right\rceil^{i'} \cr
& \qquad \qquad \times A_{R}(\alpha;\sigma_1)\frac{r+s}{2}A_{R'}(\beta;\sigma_2)
  \left\lfloor \int_a^{\sigma_2} A \right\rceil^{n-i}   \left\lfloor \int_c^{\sigma_2} A' \right\rceil^{n'-i'-1} \cr
&+(-)^{n+n'+1}  \int_{[a,b]\cap[c,d]} d\sigma
 \left\lfloor \int_{\sigma_1}^b A \right\rceil^{n-1}   \left\lfloor \int_{\sigma_1}^d A' \right\rceil^{n'} 
 A_{R}(\alpha;\sigma)\frac{r+s}{2} \cr
&-(-)^{n+n'+1} \sum_{i=1}^{n-1} \sum_{i'=0}^{n'} \int_{[a,b]\cap[c,d]} d\sigma_2 \int_{\sigma_2}^b d \sigma_1
 \left\lfloor \int_{\sigma_1}^b A \right\rceil^{i-1}   \left\lfloor \int_{\sigma_1}^d A' \right\rceil^{i'} \cr
& \qquad \qquad \times A_{R}(\alpha;\sigma_1)\frac{r+s}{2}A_{R}(\alpha;\sigma_2)
  \left\lfloor \int_a^{\sigma_2} A \right\rceil^{n-i-1}   \left\lfloor \int_c^{\sigma_2} A' \right\rceil^{n'-i'} \cr
&-(-)^{n+n'+1} \sum_{i=0}^{n-1} \sum_{i'=1}^{n'} \int_{[a,b]\cap[c,d]} d\sigma_2 \int_{\sigma_2}^d d \sigma_1
 \left\lfloor \int_{\sigma_1}^b A \right\rceil^{i}   \left\lfloor \int_{\sigma_1}^d A' \right\rceil^{i'-1} \cr
& \qquad \qquad \times A_{R'}(\beta;\sigma_1)\frac{r+s}{2}A_{R}(\alpha;\sigma_2)
  \left\lfloor \int_a^{\sigma_2} A \right\rceil^{n-i-1}   \left\lfloor \int_c^{\sigma_2} A' \right\rceil^{n'-i'} \cr
&-(-)^{n+n'+1}  \int_{[a,b]\cap[c,d]} d\sigma 
\frac{r+s}{2}A_{R}(\alpha;\sigma)
  \left\lfloor \int_a^{\sigma_2} A \right\rceil^{n-1}   \left\lfloor \int_c^{\sigma_2} A' \right\rceil^{n'} 
\end{align}
The third contribution \eqref{fusionTTO1Term3} can be rewritten in a form similar to the first two. We have to perform the integral over the derivative of the delta-function. We get a non-zero contribution only if the two connections on both sides of the constant matrix $s$ are associated to different representations:
\begin{align}
&(-)^{n+n'+2} \sum_{i=1}^{n} \sum_{i'=0}^{n'-1} \int_{a}^b d\sigma_1 \int_c^{\sigma_1} d \sigma_2
 \left\lfloor \int_{\sigma_1}^b A \right\rceil^{i-1}   \left\lfloor \int_{\sigma_1}^d A' \right\rceil^{i'} \cr
& \qquad \qquad \times A_{R}(\alpha;\sigma_1)s A_{R'}(\beta;\sigma_2)
  \left\lfloor \int_a^{\sigma_2} A \right\rceil^{n-i}   \left\lfloor \int_c^{\sigma_2} A' \right\rceil^{n'-i'-1} \cr
&-(-)^{n+n'+2} \sum_{i=0}^{n-1} \sum_{i'=1}^{n'} \int_{c}^d d \sigma_1 \int_{a}^{\sigma_1} d\sigma_2 
 \left\lfloor \int_{\sigma_1}^b A \right\rceil^{i}   \left\lfloor \int_{\sigma_1}^d A' \right\rceil^{i'-1} \cr
& \qquad \qquad \times A_{R'}(\beta;\sigma_1)s A_{R}(\alpha;\sigma_2)
  \left\lfloor \int_a^{\sigma_2} A \right\rceil^{n-i-1}   \left\lfloor \int_c^{\sigma_2} A' \right\rceil^{n'-i'} \cr
&+(-)^{n+n'+2}\chi(b;c,d) \int_{c}^d d \sigma s A_{R'}(\beta;\sigma)
  \left\lfloor \int_a^{\sigma} A \right\rceil^{n}   \left\lfloor \int_c^{\sigma} A' \right\rceil^{n'-1} \cr
&-(-)^{n+n'+2}\chi(d;a,b) \int_{a}^b d \sigma s A_{R}(\alpha;\sigma)
  \left\lfloor \int_a^{\sigma} A \right\rceil^{n-1}   \left\lfloor \int_c^{\sigma} A' \right\rceil^{n'} \cr
&+(-)^{n+n'+2} \chi(c;a,b) \int_{a}^b d\sigma 
 \left\lfloor \int_{\sigma}^b A \right\rceil^{n-1} \left\lfloor \int_{\sigma}^d A' \right\rceil^{n'} 
  A_{R}(\alpha;\sigma) s \cr
&-(-)^{n+n'+2} \chi(c;c,d) \int_{c}^d d\sigma 
 \left\lfloor \int_{\sigma}^b A \right\rceil^{n} \left\lfloor \int_{\sigma}^d A' \right\rceil^{n'-1} 
  A_{R'}(\beta;\sigma) s \cr
\end{align}
Now we sum the three contributions.
First let us consider the terms involving the matrix $r$ only. Using formula \eqref{IcapI-IIcap} we obtain:
\begin{align}
(-)^{n+n'}& \left( \chi(b;c,d) \sum_{i'=0}^{n'} \left\lfloor \int_{b}^d A' \right\rceil^{i'} \frac{r}{2} \left\lfloor \int_{a}^b A \right\rceil^{n} \left\lfloor \int_{c}^b A' \right\rceil^{n'-i'}\right. \cr
&-\chi(a;c,d) \sum_{i'=0}^{n'}  \left\lfloor \int_{a}^b A \right\rceil^{n} \left\lfloor \int_{a}^d A' \right\rceil^{i'}\frac{r}{2}  \left\lfloor \int_{c}^a A' \right\rceil^{n'-i'} \cr
&+\chi(d;a,b) \sum_{i=0}^{n} \left\lfloor \int_{d}^b A \right\rceil^{i} \frac{r}{2} \left\lfloor \int_{a}^d A \right\rceil^{n-i} \left\lfloor \int_{c}^d A' \right\rceil^{n'} \cr
&\left. -\chi(c;a,b) \sum_{i=0}^{n'}\left\lfloor \int_{c}^b A \right\rceil^{i} \left\lfloor \int_{c}^d A' \right\rceil^{n'} \frac{r}{2} \left\lfloor \int_{a}^c A \right\rceil^{n-i}  \right)
\end{align}
The terms involving the matrix $s$ simplify thanks to equation \eqref{IforS}:
\begin{align}
(-)^{n+n'}& \left( \chi(b;c,d) \sum_{i'=0}^{n'} \left\lfloor \int_{b}^d A' \right\rceil^{i'} \frac{s}{2} \left\lfloor \int_{a}^b A \right\rceil^{n} \left\lfloor \int_{c}^b A' \right\rceil^{n'-i'}\right. \cr
&-\chi(a;c,d) \sum_{i'=0}^{n'}  \left\lfloor \int_{a}^b A \right\rceil^{n} \left\lfloor \int_{a}^d A' \right\rceil^{i'}\frac{s}{2}  \left\lfloor \int_{c}^a A' \right\rceil^{n'-i'} \cr
&+\chi(d;a,b) \sum_{i=0}^{n} \left\lfloor \int_{d}^b A \right\rceil^{i} \frac{-s}{2} \left\lfloor \int_{a}^d A \right\rceil^{n-i} \left\lfloor \int_{c}^d A' \right\rceil^{n'} \cr
&\left. -\chi(c;a,b) \sum_{i=0}^{n'}\left\lfloor \int_{c}^b A \right\rceil^{i} \left\lfloor \int_{c}^d A' \right\rceil^{n'} \frac{-s}{2} \left\lfloor \int_{a}^c A \right\rceil^{n-i}  \right)
\end{align}
Thus we obtain formula \eqref{fusionf2nn'}.


\subsection{Second-order computations}\label{CompFusionO2}

The simplifications occurring in the second-order computation leading to \eqref{fusionf4nn'} are almost identical to the previous case.
The only subtlety comes from the new terms in \eqref{fusionTTO2Term3}. 
These terms come from the OPE  between two connection separated by a constant matrix $r\pm s$.
Here we explain how to deal with these terms.
The first and second terms in \eqref{fusionTTO2Term3} can be rewritten as:
\begin{align}
& (-)^{n+n'+2}  \sum_{i'=0}^{n'} \left(
 \int_a^b d\sigma \int_b^d d\sigma'  \left\lfloor \int_{\sigma'}^d A' \right\rceil^{i'} \frac{r+s}{2} s \delta'_\epsilon(\sigma-\sigma')
\left\lfloor \int_{a}^\sigma A \right\rceil^{n} \left\lfloor \int_{c}^b A' \right\rceil^{n'-i'} \right. \cr
& +  \left\lfloor \int_{b}^d A' \right\rceil^{i'} \frac{r+s}{2} 
\sum_{j=1}^{n} \sum_{j'=0}^{n'-1-i'}
\int_a^b d \sigma_1 \int_c^{\sigma_1} d \sigma_2
\left\lfloor \int_{\sigma_1}^b A \right\rceil^{j-1} \left\lfloor \int_{\sigma_1}^b A' \right\rceil^{j'} \cr
& \times A_{R}(\alpha;\sigma_1) s A_{R'}(\beta;\sigma_2)
\left\lfloor \int_{a}^{\sigma_2} A \right\rceil^{n-j} \left\lfloor \int_{c}^{\sigma_2} A' \right\rceil^{n'-i'-j'-1} \cr
& +  \left\lfloor \int_{b}^d A' \right\rceil^{i'} \frac{r+s}{2} 
\sum_{j=0}^{n-1} \sum_{j'=1}^{n'-i'}
\int_c^b d \sigma_1 \int_a^{\sigma_1} d \sigma_2
\left\lfloor \int_{\sigma_1}^b A \right\rceil^{j} \left\lfloor \int_{\sigma_1}^b A' \right\rceil^{j'-1} \cr
& \times A_{R'}(\beta;\sigma_1) s A_{R}(\alpha;\sigma_2)
\left\lfloor \int_{a}^{\sigma_2} A \right\rceil^{n-j-1} \left\lfloor \int_{c}^{\sigma_2} A' \right\rceil^{n'-i'-j'} \cr
& \left. +  \left\lfloor \int_{b}^d A' \right\rceil^{i'} \frac{r+s}{2} 
\int_a^b d \sigma \int_c^{b} d \sigma' s \delta'_{\epsilon}(\sigma-\sigma')
\left\lfloor \int_{a}^{\sigma_2} A \right\rceil^{n} \left\lfloor \int_{c}^{\sigma_2} A' \right\rceil^{n'-i'}
 \right)
\end{align}
We notice that the first and last term combine to give:
\begin{align}
& (-)^{n+n'+2}  \sum_{i'=0}^{n'} \left(
 \left\lfloor \int_{b}^d A' \right\rceil^{i'} \frac{r+s}{2} 
\sum_{j=1}^{n} \sum_{j'=0}^{n'-1-i'}
\int_a^b d \sigma_1 \int_c^{\sigma_1} d \sigma_2
\left\lfloor \int_{\sigma_1}^b A \right\rceil^{j-1} \left\lfloor \int_{\sigma_1}^b A' \right\rceil^{j'} \right. \cr
& \times A_{R}(\alpha;\sigma_1) s A_{R'}(\beta;\sigma_2)
\left\lfloor \int_{a}^{\sigma_2} A \right\rceil^{n-j} \left\lfloor \int_{c}^{\sigma_2} A' \right\rceil^{n'-i'-j'-1} \cr
& +  \left\lfloor \int_{b}^d A' \right\rceil^{i'} \frac{r+s}{2} 
\sum_{j=0}^{n-1} \sum_{j'=1}^{n'-i'}
\int_c^b d \sigma_1 \int_a^{\sigma_1} d \sigma_2
\left\lfloor \int_{\sigma_1}^b A \right\rceil^{j} \left\lfloor \int_{\sigma_1}^b A' \right\rceil^{j'-1} \cr
& \times A_{R'}(\beta;\sigma_1) s A_{R}(\alpha;\sigma_2)
\left\lfloor \int_{a}^{\sigma_2} A \right\rceil^{n-j-1} \left\lfloor \int_{c}^{\sigma_2} A' \right\rceil^{n'-i'-j'} \cr
& \left. +  \left\lfloor \int_{b}^d A' \right\rceil^{i'} \frac{r+s}{2} 
 \int_c^{b} d \sigma s A_{R'}(\beta;\sigma)
\left\lfloor \int_{a}^{\sigma} A \right\rceil^{n} \left\lfloor \int_{c}^{\sigma} A' \right\rceil^{n'-i'-1}
 \right)
\end{align}
So essentially we are back to the first-order computation.
A similar simplification happens for the last two terms in \eqref{fusionTTO2Term3}.


\subsection{Integrations over intersection of intervals}\label{intCap}

The following formulas are useful in the computation of the fusion of line operators.
The integral over an intersection of intervals can be rewritten as:
\begin{align}\label{cap=3chi} \int_{[a,b]\cap[c,d]} dz f(z) 
&= \chi(a;c,d) \int_a^d dz f(z) + \chi(c;a,b) \int_c^d dz f(z) - \chi(b;c,d) \int_b^d dz f(z) \cr
&= \chi(a;c,d) \int_a^b dz f(z) + \chi(c;a,b) \int_c^b dz f(z) - \chi(d;a,b) \int_d^b dz f(z) \cr
&= \chi(d;a,b) \int_a^d dz f(z) + \chi(b;c,d) \int_a^b dz f(z) - \chi(c;a,b) \int_a^c dz f(z) \cr
&= \chi(d;a,b) \int_c^d dz f(z) + \chi(b;c,d) \int_c^b dz f(z) - \chi(a;c,d) \int_c^a dz f(z) \end{align}
This can be generalized as follows:
\begin{align}\label{IcapI-IIcap} &\int_{[a,b]\cap [c,d]} dz_1 \int_{[a,z_1]} dz_2 f(z_1,z_2) - \int_{[z_2,b]} dz_1 \int_{[a,b]\cap [c,d]} dz_2  f(z_1,z_2) \cr &= \chi(c;a,b)\int_c^b dz_1 \int_a^c dz_2 f(z_1,z_2) - \chi(d;a,b)\int_d^b dz_1 \int_a^d dz_2 f(z_1,z_2) \end{align}
%
%
The following formula also plays a r\^ole:
\begin{align}\label{IforS}  \int_{\genfrac{}{}{0pt}{}{\sigma' \in [c,d]}{\sigma \in[a,b],\ \sigma'>\sigma}} f(\sigma',\sigma) 
& = \chi(b;c,d) \int_b^d d\sigma' \int_a^b d\sigma f(\sigma',\sigma) + \int_{\genfrac{}{}{0pt}{}{\sigma \in [a,b],\ \sigma'>\sigma}{\sigma'\in[a,b]\cap[c,d]}} f(\sigma',\sigma) \cr
& = \chi(c;a,b) \int_c^d d\sigma' \int_a^c d\sigma f(\sigma',\sigma) + \int_{\genfrac{}{}{0pt}{}{\sigma' \in [c,d],\ \sigma'>\sigma}{\sigma\in[a,b]\cap[c,d]}} f(\sigma',\sigma)
\end{align}


\section{Character identities}\label{charIdentities}

Consider a group element $g \in Gl(k|m)$ in a representation associated to a rectangular Young tableau labeled by two integers $(a,s)$. 
The associated supercharacters satisfy (see e.g. \cite{Kazakov:2007na}):
\be\label{charId(1)'} \chi(a,s)^2 = \chi(a+1,s)\chi(a-1,s)+\chi(a,s+1)\chi(a,s-1) \ee
This is the classical version of the Hirota equation.
In \cite{Kazakov:2007na} the quantum Hirota equation was proven for transfer matrices associated to $Gl(k|m)$ spin chains. 
Various character identities can be deduced.
In particular we have for any group element and for any integer $N$:
\begin{align}\label{charId(N)'} &(1+2\hat D)^{\otimes N}\chi(a,s) (-1+2\hat D)^{\otimes N}\chi(a,s) \cr
& = (1+2\hat D)^{\otimes N}\chi(a+1,s) (-1+2\hat D)^{\otimes N}\chi(a-1,s)\cr
&+ (1+2\hat D)^{\otimes N}\chi(a,s-1) (-1+2\hat D)^{\otimes N}\chi(a,s+1)
\end{align}
The quantities appearing on both sides of the previous equation are linear operators acting on the $N$-times tensor product of the fundamental representation of $Gl(k|m)$.
We can think of these operators as acting on a spin chain with $N$ sites, with a spin in the fundamental representation at each site.
The operator $\hat D$ acts on a supercharacter as:
\be \hat D^i_j STr_{R}(g) = (-)^j STr_{R}(t_{ji} g) \ee
where $i,j...$ are indices in the fundamental representation of $Gl(k|m)$ and $t_{ij}$ is a generator. 
The sign $(-)^j$ is $(+)$ if $1\le j \le k$, and $(-)$ if $k+1\le j \le k+m$.

Using the function $\H$ defined in \eqref{defH}, the Jacobi-Trudi identity \eqref{charId(1)'} is conveniently written as:
\be\label{charId(1)} \forall g \in Gl(k|m), \quad \H(g\otimes g) = 0 \ee
Similarly, equation \eqref{charId(N)'} is written as:
\begin{align}\label{charId(N)} \forall g \in Gl(k|m), \quad \sum_{j_1,...,j_N} \H & \left((1+2(-)^{i_1}t_{i_1 j_1})...(1+2(-)^{i_N}t_{i_N j_N})g \right. \cr 
& \qquad \left.\otimes (-1+2(-)^{j_1}t_{j_1 k_1})...(-1+2(-)^{j_N}t_{j_N k_N})g  \right) = 0 \end{align}
Next we consider this identity in some particular cases. 
In this appendix we keep track of the additional signs that appears in the computations due to the fermionic natures of some of the generators.
We adopt the following conventions (``South-West North-East'') for the contraction of super-indices:
\begin{align}
&\kappa_{ij,kl} \kappa^{mn,kl} = \delta^m_i \delta^n_j \cr
& A_a = \kappa_{ab} A^b \qquad ; \qquad A^a = A_b \kappa^{ba} \cr
&[t_a,t_b] = i t_c {f^c}_{ab}
\end{align}
%
Let us introduce some data for the group $Gl(k|m)$.
We consider a basis of generators $t_{ij}$. In the fundamental representation, the generators read explicitly:
\be (t_{ij})^{k}_l = \delta_i^k \delta_{j,l} \ee
The metric is:
\be \kappa_{ij,kl} = STr(t_{ij} t_{kl}) = (-)^i \delta_{i,l} \delta_{j,k}  
\qquad ; \qquad
\kappa^{ij,kl} = (-)^i \delta_{i,l} \delta_{j,k} \ee
In particular:
\be A_{ij} B_{kl} \kappa^{kl,ij} = \sum_{i,j} A_{ij} B_{ji} (-)^j \ee
The structure constants read:
%
%
%
\be {f^{mn}}_{ij,kl} = \delta_{j,k}\delta_{m,i} \delta_{n,l} - (-)^{(i+j)(k+l)} \delta_{i,l}\delta_{m,k} \delta_{j,n} \ee
%
%
%
Notice also that:
\be \sum_i t_{ii} = 1 \ee

Now let us consider the relation \eqref{charId(N)} with $N=1$. Using \eqref{charId(1)} it can be rewritten as:
\begin{align}\label{charId(2)}
2\sum_j \H(t_{ij}g\otimes t_{jk} g)(-)^{i+j} = (-)^i \H(t_{ik}g\otimes g) - (-)^i \H(g\otimes t_{ik}g)
\end{align}
We set $i=k$ in equation \eqref{charId(2)}. Then we multiply by $(-)^i$ and sum over $i$. We obtain:
\be 2\sum_{i,j} \H(t_{ji}g\otimes t_{ij} g) (-)^j = \sum_i \H(t_{ii}g\otimes g) - \H(g\otimes t_{ii}g) = 0 \ee
This is conveniently written as:
\be\label{charId(2)K}  \H(t_a g\otimes  t_b g) \kappa^{ba} = 0 \ee
where $a,b$ are adjoint indices.
Now we contract the indices $i,k$ in equation \eqref{charId(2)} with an arbitrary function $J^{ki}$ and multiply by $(-)^i$. We obtain (using \eqref{charId(1)}):
\be\label{charId(2)J} 2 \sum_{i,j,k} \H(t_{ij}g\otimes t_{jk} g) (-)^j J^{ki} = \sum_{i,k} \H(t_{ik}g\otimes g)J^{ki} - \sum_{i,k} \H(g\otimes t_{ik}g)J^{ki} \ee
Let us consider the following product of structure constants and generators contracted with an arbitrary function $J^{ij}$ carrying two fundamental indices:
\begin{align} t^R_{op} t^R_{ij} t^{R'}_{mn} i{f^{mn,ij}}_{kl}i{f^{kl,op}}_{qr}  J^{qr} \end{align}
where $R$ and $R'$ label two representations.
Using the explicit expression for the structure constants, together with:
\be t_{ij} t_{kl} = t_{il} \delta_{jk} \ee
we obtain for the previous quantity:
\begin{align}  &
(-)^i \delta_{ii} t^R_{qm} t^{R'}_{mr}(-)^m J^{qr}
-t^R_{nm} t^{R'}_{mn}(-)^m \delta_{qr}J^{qr}(-)^r
-t^R_{qr}t^{R'}_{nn}J^{qr} 
+ t^{R}_{jj} t^{R'}_{qr}J^{qr}  \cr
& = (k-m) t^R_{qm} t^{R'}_{mr}(-)^mJ^{qr} 
-t^R_{nm} t^{R'}_{mn}(-)^m \delta_{qr}J^{qr}(-)^r
-t^R_{qr}J^{qr} 
+  t^{R'}_{qr}J^{qr} 
\end{align}
where a sum over all repeated indices is implicit.
We deduce, using \eqref{charId(2)K} and \eqref{charId(2)J}:
\begin{align}  & \H( t_{op} t_{ij}g\otimes  t_{mn}g)  i{f^{mn,ij}}_{kl}i{f^{kl,op}}_{qr}  J^{qr}\cr
&= (k-m) \H(t_{qm}g\otimes t_{mr} g)(-)^m J^{qr} + \H(g\otimes t_{ij}g)J^{ij} - H(t_{ij}g,g) J^{ij} \cr
&= \left(1-\frac{k-m}{2}\right)(\H(t_{ij}g\otimes g) J^{ij} - \H(g\otimes t_{ij}g)J^{ij})
\end{align}
This is rewritten in a more convenient way using adjoint indices $a,b,c...$:
\begin{align}\label{charId(2)ffJ}
\H(t_{a} t_{b} g\otimes  t_c g) i{f^{cb}}_d i{f^{da}}_e J^e = \left(1-\frac{k-m}{2}\right) (-\H( t_a J^a g\otimes g)+\H(g\otimes  t_a J^a g))
\end{align}%
Similarly we can show that:
\begin{align}\label{charId(2)ffJ'}
\H(t_{a} g\otimes t_{b} t_{c} g) i{f^{c}}_{de} J^e i{f^{dba}} = \left(1-\frac{k-m}{2}\right) (\H( t_a J^a g\otimes g)-\H(g\otimes  t_a J^a g))
\end{align}


\subsection*{Generalization to $PSl(n|n)$}

In the previous paragraph several characters identities were derived for the supergroup $Gl(k|m)$. Now we will show that some of these identities are also valid for the supergroup $PSl(n|n)$.

The supergroup $PSl(n|n)$ is obtained as the quotient of the supergroup $Gl(n|n)$ by the action of two $U(1)$ generators: the identity $I$ and the generator $\tilde{I}$ such that $\tilde{I}^i_j = \delta^i_j$ for $1 \le i \le n$, and $\tilde{I}^i_j = -\delta^i_j$ for $n+1 \le i \le 2n$.

First the identity \eqref{charId(1)} generalizes trivially, by taking a group element $g$ into $PSl(n|n)$:
\be\label{charId(1)PSL} \forall g \in PSl(n|n), \qquad \H(g\otimes g) = 0 \ee
To generalize the identity \eqref{charId(2)K}, we split the $gl(n|n)$ generators as $psl(n|n)$ generators plus $I$ and $\tilde{I}$. We notice that the generators $I$ and $\tilde{I}$ belong to the supergroup $Gl(n|n)$. Then using \eqref{charId(1)} we obtain:
\be\label{charId(2)KPSL} \forall g \in PSl(n|n), \qquad  \H(t_a g\otimes t_b g)\kappa^{ba} = 0 \ee
where $a,b$ are now indices in the adjoint representation of $PSl(n|n)$. Next the identities \eqref{charId(2)ffJ} and \eqref{charId(2)ffJ'} also generalizes since the generators $I$ and $\tilde{I}$ commute with all other generators:
\begin{align}\label{charId(2)ffJPSL}
\forall g \in PSl(n|n), \quad  
& \H(t_{a} t_{b} g\otimes  t_c g) i{f^{cb}}_d i{f^{da}}_e J^e =  -\H( t_a J^a g\otimes g)+\H(g\otimes  t_a J^a g) \cr
& \H(t_{a} g\otimes t_{b} t_{c} g) i{f^{c}}_{de} J^e i{f^{dba}} =  \H( t_a J^a g\otimes g)-\H(g\otimes  t_a J^a g)
\end{align}

\end{appendix}

\end{document}